\documentclass[aps,11pt,final,preprint,nobibnotes,nofootinbib,superscriptaddress,centertags,preprintnumber]{revtex4}
 \usepackage[dvips,final]{graphicx}
  \usepackage{pstricks,amsmath,amssymb,pifont}

\graphicspath{{./Figs-BMPS/}}

  \newrgbcolor{violet}{0.4 0 0.9}

\begin{document}
 \preprint{\hbox{RUB-TPII-06/07}}

\title{Pion structure in QCD: From theory to lattice to experimental data\footnote{%
Invited plenary talk presented by the first author at ``Hadron Structure '07''
International Conference, Modra-Harm{\'o}nia, Slovakia, Sept. 2--7, 2007.}}

\author{A.~P.~Bakulev}
\affiliation{Bogoliubov Lab. of Theoretical Physics, JINR, 141980 Dubna, Russia}
\email{bakulev@theor.jinr.ru}

\author{S.~V.~Mikhailov}
\affiliation{Bogoliubov Lab. of Theoretical Physics, JINR, 141980 Dubna, Russia}
\email{mikhs@theor.jinr.ru}

\author{A.~V.~Pimikov}
\affiliation{Bogoliubov Lab. of Theoretical Physics, JINR, 141980 Dubna, Russia}
\email{pimikov@theor.jinr.ru}

\author{N.~G.~Stefanis}
\affiliation{ITP-II, Ruhr-Universit\"at Bochum, D-44780 Bochum, Germany}
\email{stefanis@tp2.ruhr-uni-bochum.de}

\begin{abstract}
 We describe the present status of the pion distribution amplitude (DA)
 as it originates from several sources:
 (i) a nonperturbative approach based on QCD sum rules with nonlocal
 condensates,
 (ii) an $O(\alpha_s)$ QCD analysis of the CLEO data on
 $F^{\gamma\gamma^*\pi}(Q^2)$
 with asymptotic and renormalon models for higher twists, and
 (iii) recent high-precision lattice QCD calculations 
 of the second moment of the pion DA.
 We show predictions for the pion electromagnetic form factor,
 obtained in analytic QCD perturbation theory,
 and compare it with the JLab data on $F_{\pi}(Q^2)$.
 We also discuss in this context an improved model for nonlocal condensates 
 in QCD and show its consequences for the pion DA and 
 the $\gamma\gamma^*\to\pi$ transition form factor.
 We include a brief analysis of meson-induced massive lepton
 (muon) Drell--Yan production for the process
 $\pi^{-}N\to\mu^{+}\mu^{-}X$, considering 
 both an unpolarized nucleon target and longitudinally polarized protons.
\end{abstract}



\maketitle

\section{Introduction}
The pion DA parameterizes
the matrix element of the nonlocal axial current on the light cone~\cite{Rad77}
\begin{eqnarray}
 \label{eq:pion.DA.ME}
 \langle{0\!\mid\!\bar d(z)\gamma_{\mu}\gamma_5 
  {\cal C}(z,0)u(0)\!\mid\!\pi(P)}\rangle\Big|_{z^2=0}
  = i f_{\pi}P_{\mu}\!
       \int\limits_{0}^{1}\! dx\ e^{ix(zP)}
        \varphi_{\pi}^\text{Tw-2}(x,\mu^2)\,.~~
\end{eqnarray}
The gauge-invariance of this DA is ensured 
by the Fock--Schwinger connector~\cite{Ste84} 
(Wilson line)
$${\cal C}(z,0)={\cal P}\exp\left[i g\int_0^z A_\mu(\tau) d\tau^\mu\right],$$
inserted between the two quark fields.
The physical meaning of this DA is quite evident: 
it is the amplitude 
for the transition $\pi(P)\rightarrow u(Px) + \bar{d}(P(1-x))$.
It is convenient to represent the pion DA using an expansion 
in terms of Gegenbauer polynomials $C^{3/2}_n(2x-1)$,
which are one-loop eigenfunctions of the ERBL kernel~\cite{ER80,LB79},
i.~e.,
\begin{eqnarray}
 \label{eq:pi.DA.rep}
  \varphi_\pi(x;\mu^2) 
  = \varphi^\text{As}(x)\,
     \Bigl[1 + \sum\limits_{n\geq1}a_{2n}(\mu^2)\,C^{3/2}_{2n}(2x-1)
     \Bigr]\,,
\end{eqnarray}
where $\varphi^\text{As}(x)=6\,x\,(1-x)$ is the asymptotic pion DA.
This representation means 
that all scale dependence in $\varphi_\pi(x;\mu^2)$
is transformed into the scale dependence 
of the set 
$\left\{a_2(\mu^2), a_4(\mu^2), \ldots\right\}$.
We mention here 
that the ERBL solution at the 2-loop level
is also possible 
using the same representation 
(\ref{eq:pi.DA.rep})~\cite{MR86ev,KMR86,Mul94,BS05}.

In order to construct reliable QCD SRs for the pion DA
moments,
one needs, as has been shown in~\cite{MR86,BM98},
to take into account 
the nonlocality of the QCD vacuum condensates.
For an illustration of the nonlocal condensate (NLC) model,
we use here the minimal Gaussian model
\begin{eqnarray}
 \label{eq:Min.Gauss.Mod}
  \langle{\bar{q}(0)q(z)}\rangle 
   = \langle{\bar{q}\,q}\rangle\,
      e^{-|z^2|\lambda_q^2/8}
\end{eqnarray} 
with a single scale parameter $\lambda_q^2 = \langle{k^2}\rangle$
that characterizes 
the average momentum of quarks in the QCD vacuum.
Its value has been estimated in the QCD SR approach 
and also on the lattice~\cite{BI82lam,OPiv88,DDM99,BM02}:
\begin{eqnarray}
 \label{eq:lambda.q.SR}
  \lambda_q^2
   = 0.35-0.55~\text{GeV}^{2}\,.
\end{eqnarray}  
Let us write down, as an example, 
the NLC QCD SR for the pion DA $\varphi_{\pi}(x)$.
To derive it, 
one starts from a correlator of the currents
$J_{\mu5}(x)$ and 
$J^{\dagger}_{\nu5;N}(0)=\bar{d}(0)\,\hat{n}\,\gamma_5\left(n\nabla\right)^N\!u(0)$
with a light-like vector $n$, $n^2=0$
to obtain next SRs for the moments $\langle{x^N}\rangle_{\pi}$,
and finally to apply the inverse Mellin transform
and arrive at
$\langle{x^N}\rangle_{\pi} \Rightarrow \varphi_\pi(x)$.
As a result, we then find
\begin{eqnarray}
 \label{eq:NLC.SR.pion.DA}
   f_{\pi}^2\,\varphi_\pi(x) 
   = \int_{0}^{s_{0}}\!\!\rho^\text{pert}(x;s)\,
         e^{-s/M^2}ds 
     + \frac{\alpha_s\langle GG\rangle}{24\pi M^2}\,
        \varphi_{GG}(x;\Delta)
     + \frac{8\pi\alpha_s\langle{\bar{q}q}\rangle^2}{81M^4}
        \sum_{i=2V,3L,4Q}\varphi_i(x;\Delta)
\end{eqnarray}
with $\Delta\equiv\lambda_q^2/M^2$. 
The local limit $\Delta\to0$ of this SR 
is specified by the appearance of $\delta$-functions 
concentrated at the end-points $x=0$ and $x=1$,
for example,
$\varphi_{4Q}(x;\Delta)=9[\delta(x)+\delta(1-x)]$.

The minimal Gaussian model (\ref{eq:Min.Gauss.Mod})
generates the contribution $\varphi_\text{4Q}(x;\Delta)$,
shown on the right panel of Fig.\ \ref{fig:piDA.SR.4Q}
in comparison with the perturbative one
for the standard (local) and the NLC types of that SR.
We see that due to the completely different behavior 
of the perturbative and condensate terms 
in the local QCD SR,
it is difficult to reach a reasonable consistency.
In contrast, the NLC contribution behaves similar 
to the perturbative one.
Just for this reason, 
we have a very good stability in the NLC SR case.
After processing SR (\ref{eq:NLC.SR.pion.DA}) 
for the moments $\langle{\xi^N}\rangle_\pi = \int_0^1\varphi_\pi(x)\,\left(2x-1\right)^N dx$,
we can restore the pion DA $\varphi_\pi(x)$ 
by demanding that it should reproduce the first five moments
$\langle{\xi^i}\rangle_\pi$, $i=2\,, 4\,, \ldots\/, 10$,
using to this purpose 
the minimally possible number of Gegenbauer harmonics
in representation (\ref{eq:pi.DA.rep}).
It comes out that the NLC SRs for the pion DA
yield a bunch of self-consistent two-parameter models
at $\mu_0^2\simeq 1.35$ GeV$^2$:
\begin{figure}[t]
 \centerline{\includegraphics[width=0.31\textwidth]{
     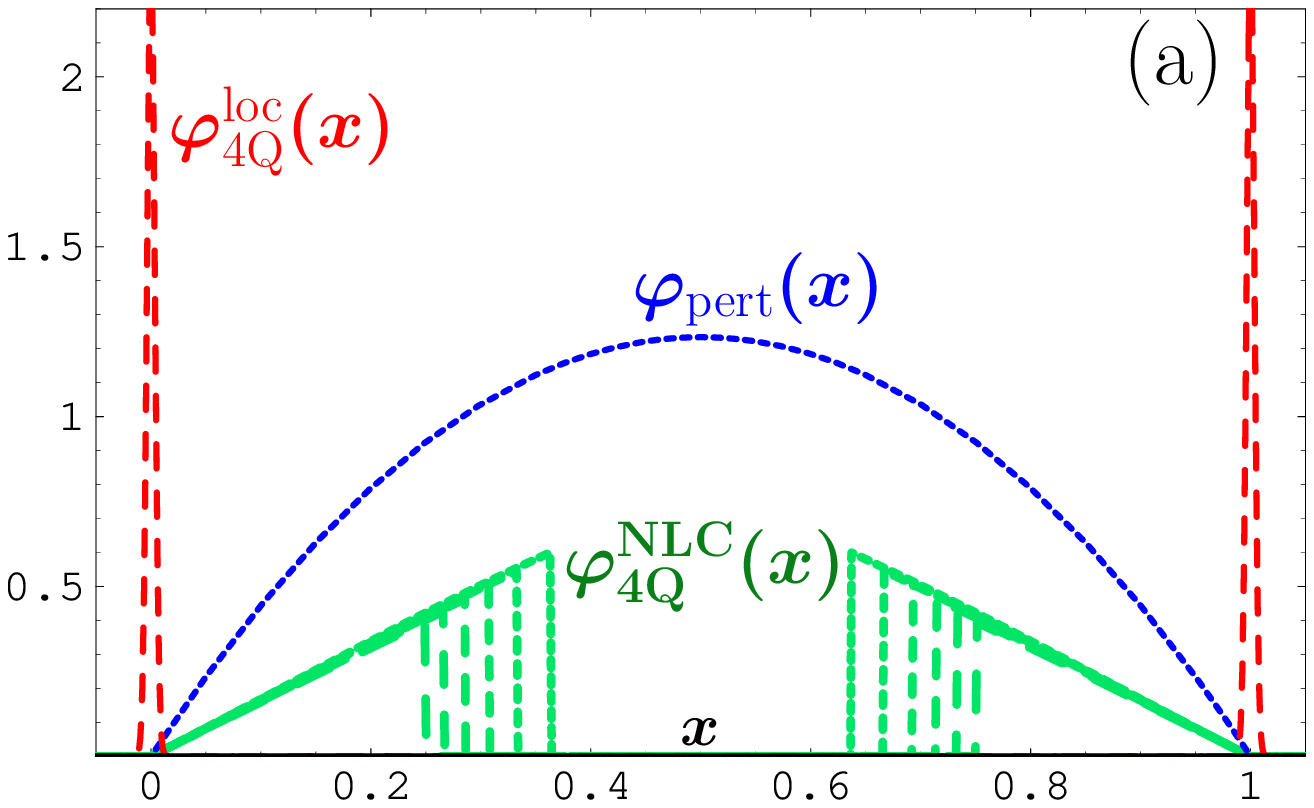}~~~
   \includegraphics[width=0.31\textwidth]{
     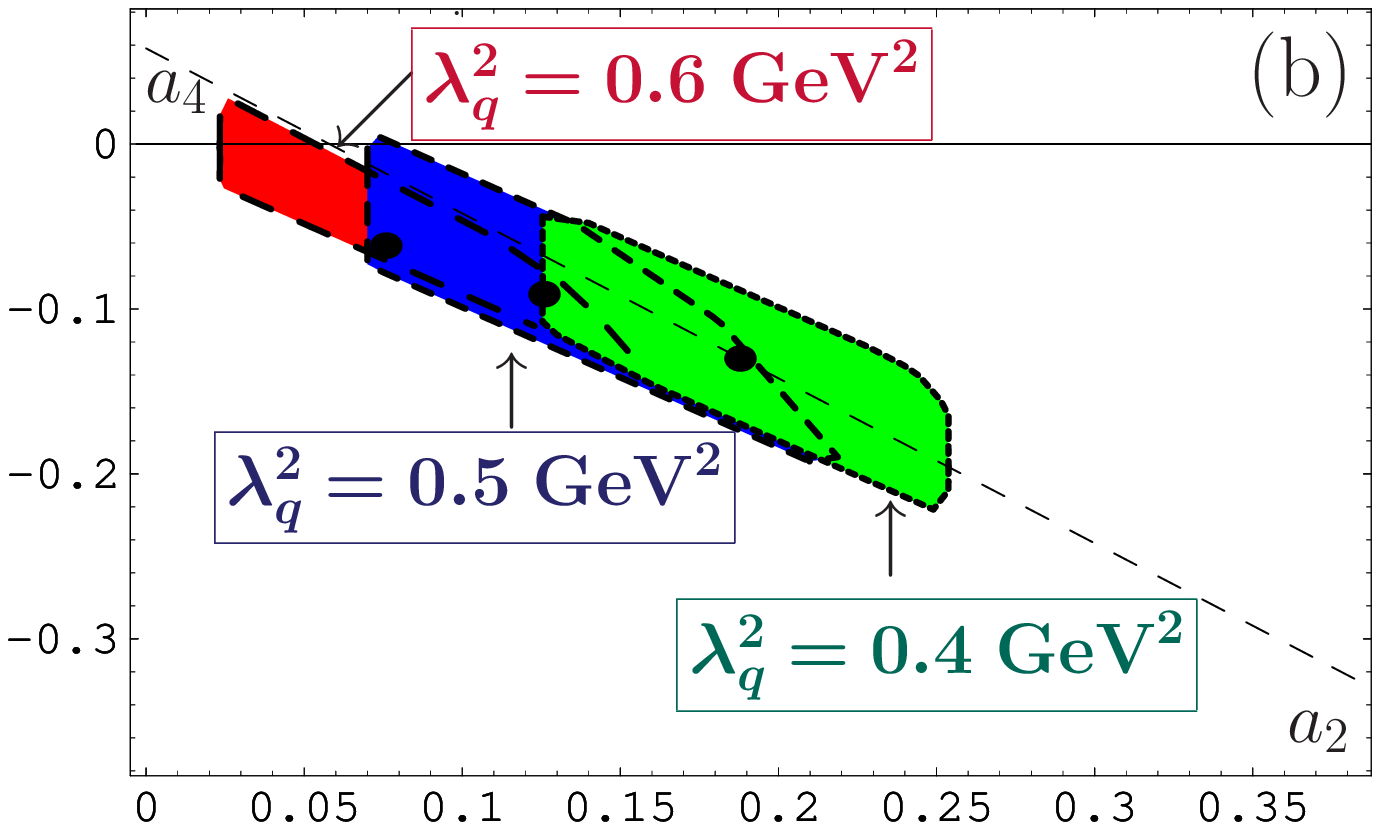}~~~
    \includegraphics[width=0.31\textwidth]{
     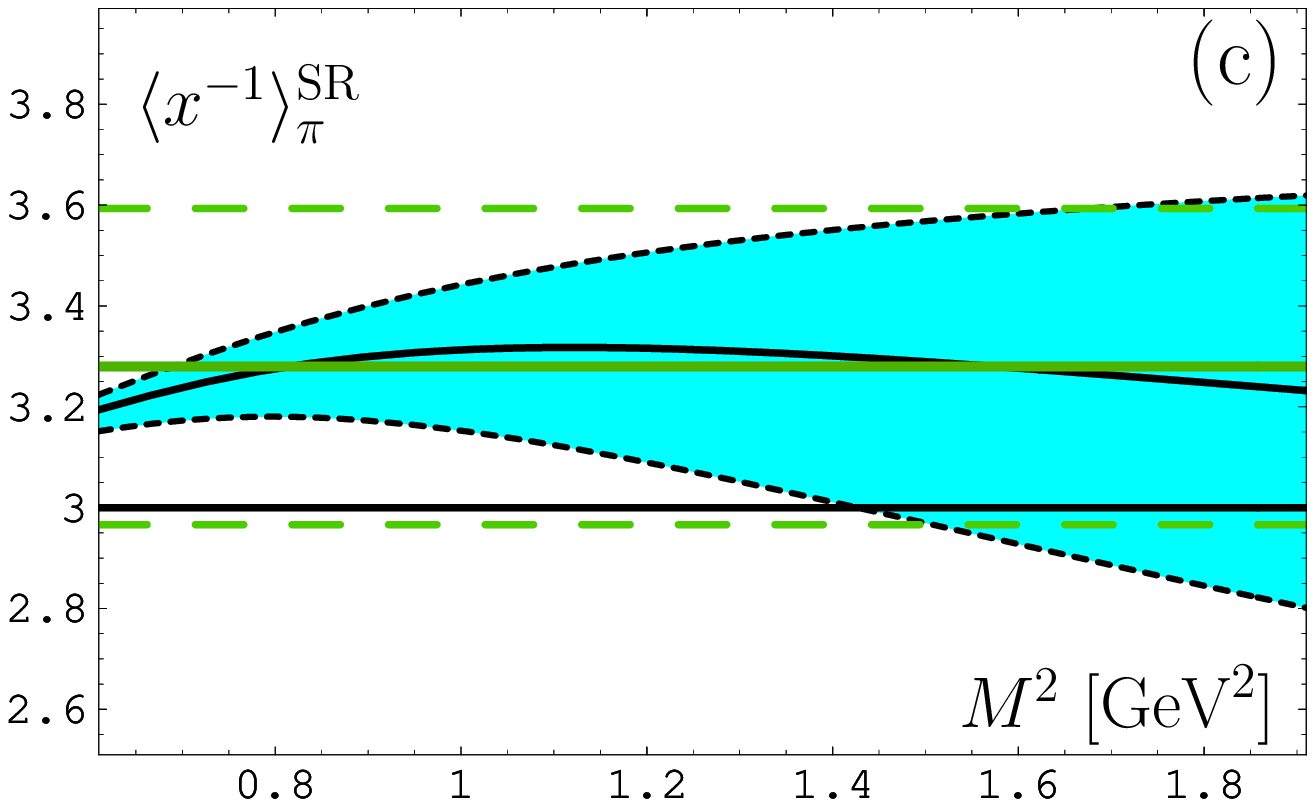}\vspace*{-3mm}}
   \caption{Panel (a): The contributions to Eq.\ (\ref{eq:NLC.SR.pion.DA})
   are shown due to the perturbative loop (dotted line) and the four-quark condensate: 
   $\varphi^\text{loc}_\text{4Q}(x)$ (standard QCD SRs);
   $\varphi^\text{NLC}_\text{4Q}(x,M^2=0.55-0.80~\text{GeV}^2)$ (the NLC QCD SRs).
  Panel (b): Allowed values of the parameters $a_2$ and $a_4$ 
  of the bunches (\ref{eq:pi.DA.2Geg}),
  evaluated at $\mu^2=1.35$~GeV$^2$, for three values 
  of the nonlocality parameter: $\lambda_q^2=0.4\,, 0.5$, and $0.6$~GeV$^2$.
  Panel (c): The inverse moment $\langle{x^{-1}}\rangle_\pi$, 
  obtained using the NLC SR (\ref{eq:NLC.SR.pion.DA}), is shown 
  as a solid line (central value) with error-bars represented by dashed lines.
  \label{fig:piDA.SR.4Q}\vspace*{-1mm}}
\end{figure}
\begin{eqnarray}
 \label{eq:pi.DA.2Geg}
  \varphi^\text{NLC}_\pi(x;\mu_0^2) 
  = \varphi^\text{As}(x)\,
     \Bigl[1 + a_{2}(\mu_0^2)\,C^{3/2}_{2}(2x-1)
             + a_{4}(\mu_0^2)\,C^{3/2}_{4}(2x-1)
     \Bigr]\,.~
\end{eqnarray}
The central point corresponds to $a_2^\text{BMS}=+ 0.188$, $a_4^\text{BMS}=-0.130$
for $\lambda^2_q=0.4$ GeV$^2$,
whereas other allowed values of the parameters $a_2$ and $a_4$
are shown on the central panel of Fig.\ \ref{fig:piDA.SR.4Q} 
as a slanted rectangle~\cite{BMS01}.
Because the inverse moments of all these pion DAs
equal
\begin{eqnarray}
 \label{eq:pion.DA.Inv.Mom}
  \langle{x^{-1}}\rangle^\text{bunch}_{\pi}
   = 3.17\pm0.20\,,
\end{eqnarray}
being in good agreement
with the estimation
dictated by an independent SR for this moment,
we term this bunch self-consistent.
This SR can be obtained 
through the basic SR (\ref{eq:NLC.SR.pion.DA})
by integrating over $x$ and using the weight $x^{-1}$ 
(at $\mu_0^2\simeq 1.35$ GeV$^2$) to find
\begin{eqnarray}
 \label{eq:Inv.Mom.SR}
 \langle{x^{-1}}\rangle_{\pi}^{\text{SR}}=3.30\pm0.30\,,
\end{eqnarray}
shown graphically on the panel (c) of Fig.\ \ref{fig:piDA.SR.4Q}.

It is worth emphasizing that the moment 
$\langle{x^{-1}}\rangle^\text{SR}_{\pi}$ could be determined 
only with NLC SRs by virtue of the absence of end-point singularities.
\begin{figure}[t]
 \centerline{\includegraphics[width=0.4\textwidth]{
    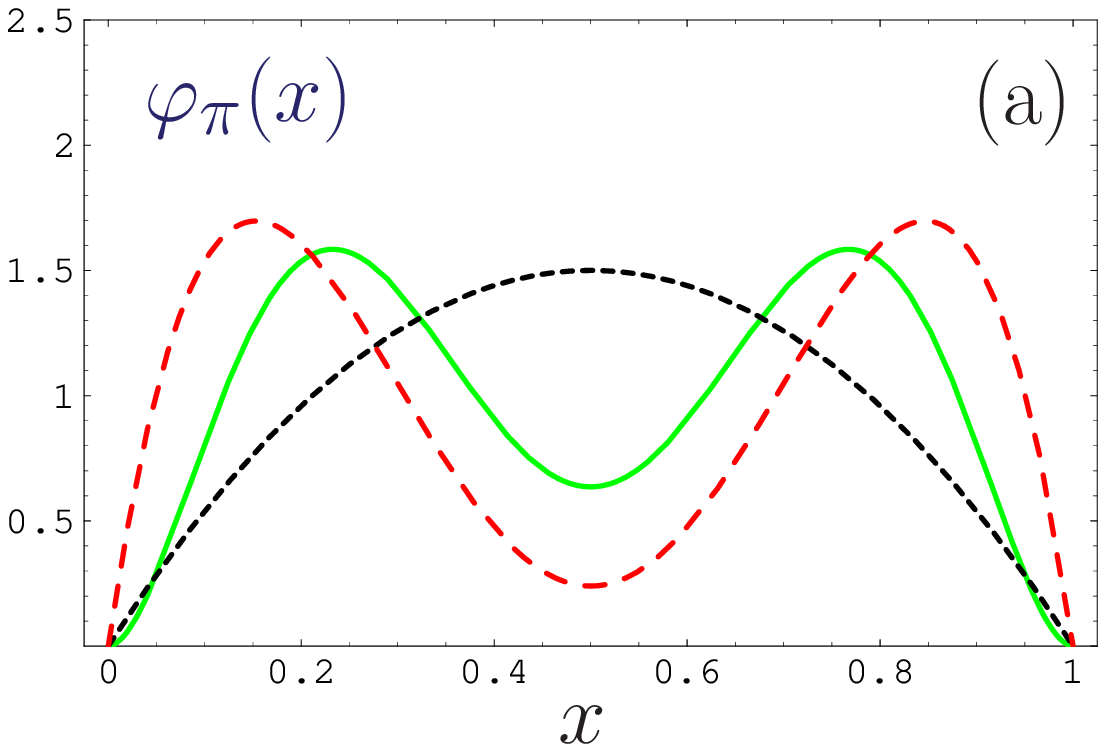}~~~~~~~~
             \includegraphics[width=0.4\textwidth]{
    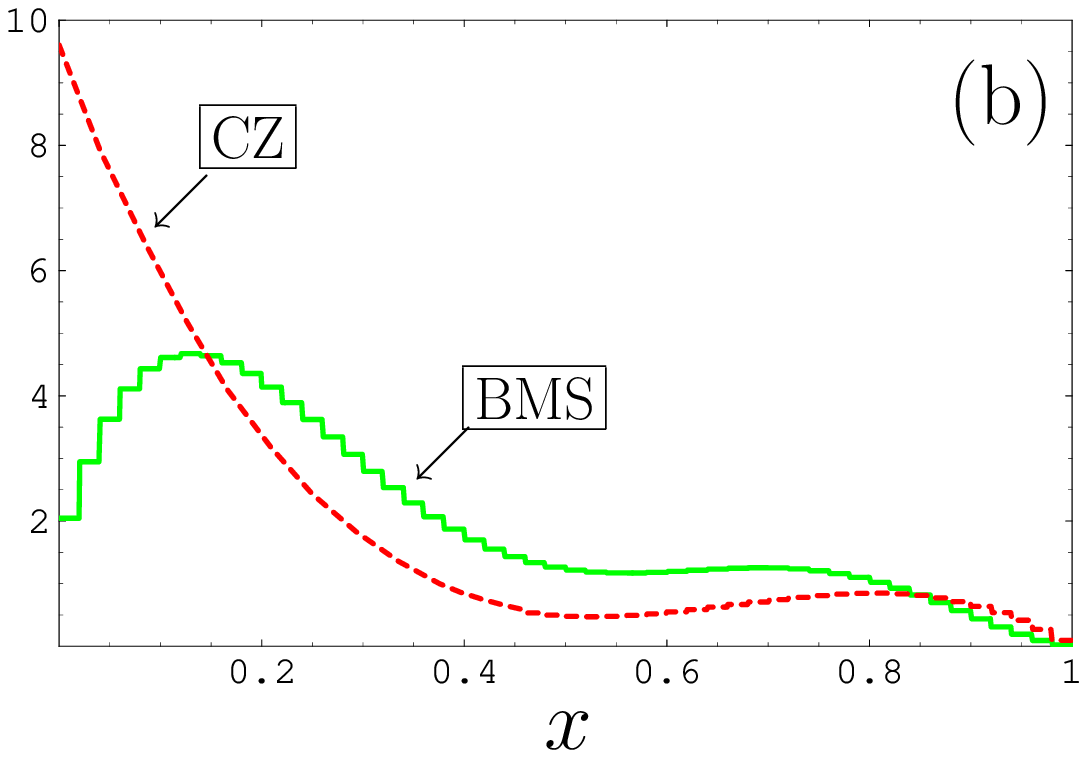}\vspace*{-3mm}}
 \caption{Panel (a): Comparison of the profiles of three pion DAs:
  BMS (solid line), CZ (dashed line), and the asymptotic DA (dotted line).
  Panel (b): Histograms for the contributions of different bins to the inverse
  moment $\langle{x^{-1}}\rangle_{\pi}$ shown for the CZ and the BMS DAs.
  \label{fig:CZ.As.BMS}\vspace*{-1mm}}
\end{figure}
Comparing the obtained pion DA with 
the Chernyak--Zhitnitsky (CZ) one~\cite{CZ82},
reveals that although both DAs are two-humped
they have distinct characteristics:
the BMS DA is strongly end-point suppressed,
as illustrated in Fig.\ \ref{fig:CZ.As.BMS}(a),
while the CZ one is end-point dominated.
To display this property more explicitly,
we show on panel (b) of this figure
the comparison of the BMS and the CZ contributions 
to the inverse moment $\langle{x^{-1}}\rangle_{\pi}$,
grouped in different bins and 
calculated as $\int_x^{x+0.02}u^{-1}\varphi(u)\,du$ 
(normalized to 100\%).

\section{Analysis of the CLEO data on $F_{\gamma\gamma^*\pi}(Q^2)$}
Many studies \cite{Kho99,SY99,SSK99,AriBro-02,BM02,BMS02}
have been performed in the literature 
to determine the pion DA 
using the high-precision CLEO data \cite{CLEO98}
on the pion-photon transition form factor 
$F_{\pi\gamma^{*}\gamma}(Q^2)$. 
In particular, in \cite{BMS02} we have used 
Light-Cone Sum Rules (LCSR) \cite{Kho99,SY99} 
to the next-to-leading-order accuracy 
of QCD perturbation theory 
to examine the theoretical uncertainties 
involved in the CLEO-data analysis 
in order to extract 
more reliably the first two non-trivial Gegenbauer coefficients 
$a_2$ and $a_4$, which parameterize the deviation
from the asymptotic expression $\varphi_{\pi}^\text{As}$.

Let us clarify why it is advantageous to use LCSRs 
in analyzing the experimental data 
on $\gamma^*(Q)\gamma(q)\to\pi^0$-transition 
form factor.
For $Q^2\gg m_\rho^2$ and $q^2\ll m_\rho^2$, 
QCD factorization is valid only in the leading-twist approximation,
so that the higher twists are important~\cite{RR96}.
The reason is quite clear: 
if $q^2\to0$, one has to take into account 
the interaction 
of a real photon at long distances 
of the order of $O(1/\sqrt{q^2})$.
Then, in order to account for long-distance effects
in perturbative QCD, 
one has to introduce the light-cone DA of the real photon.
Instead of doing so,
Khodjamirian~\cite{Kho99} suggested to use the LCSR approach,
which effectively accounts for the long-distances effects 
of the real photon using quark-hadron duality 
in the vector channel and a dispersion relation in $q^2$.
Schmedding and Yakovlev used this approach
in analyzing the CLEO data 
on the $\gamma^*\gamma\to\pi$ transition form factor
at the level of the next-to-leading-order (NLO) accuracy 
of the perturbative QCD part of the LCSR~\cite{SY99}.

We improved in~\cite{BMS02,BMS05lat} the NLO analysis of the CLEO data
by taking into account 
the following key elements: 
(i) the NLO evolution for both {$\varphi(x, Q^2_\text{exp})$} 
    and $\alpha_s(Q^2_\text{exp})$ was generalized to include
    heavy-quark thresholds more accurately;
(ii) a relation  between the ``nonlocality'' scale and 
     the twist-4 magnitude 
     $\delta^2_\text{Tw-4} \approx \lambda_q^2/2$ 
     was employed used to reestimate 
     $\delta^2_\text{Tw-4}=0.19 \pm 0.02$ 
     at $\lambda_q^2=0.4$ GeV$^2$;
(iii) the possibility to extract constraints on $\langle{x^{-1}}\rangle_\pi$ 
      from the CLEO data and 
      compare them with those we derived before from NLC QCD SRs~\cite{BMS01}
      was exploited.

\begin{figure}[h]
 \centerline{\includegraphics[width=0.32\textwidth]{
             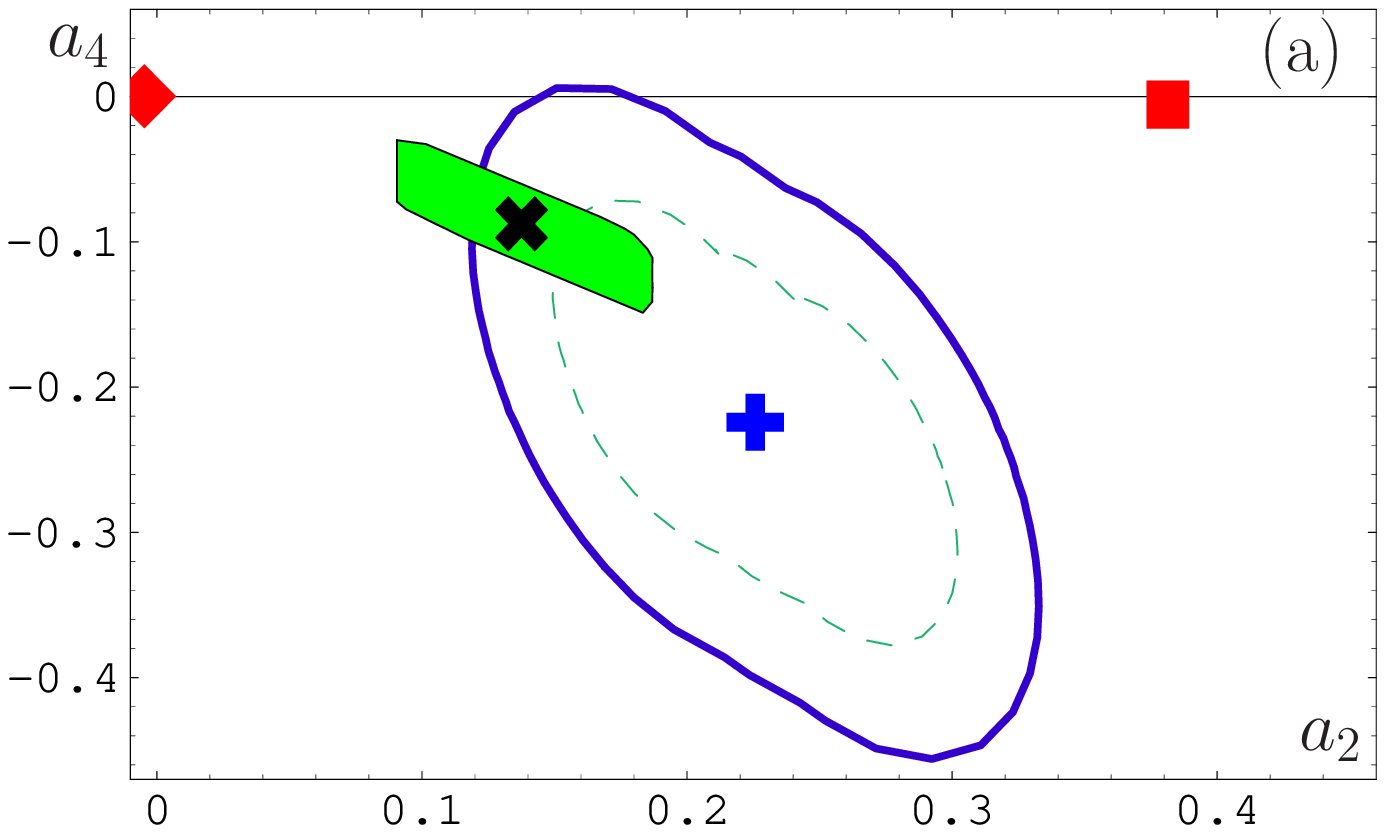}%
            ~\includegraphics[width=0.32\textwidth]{
             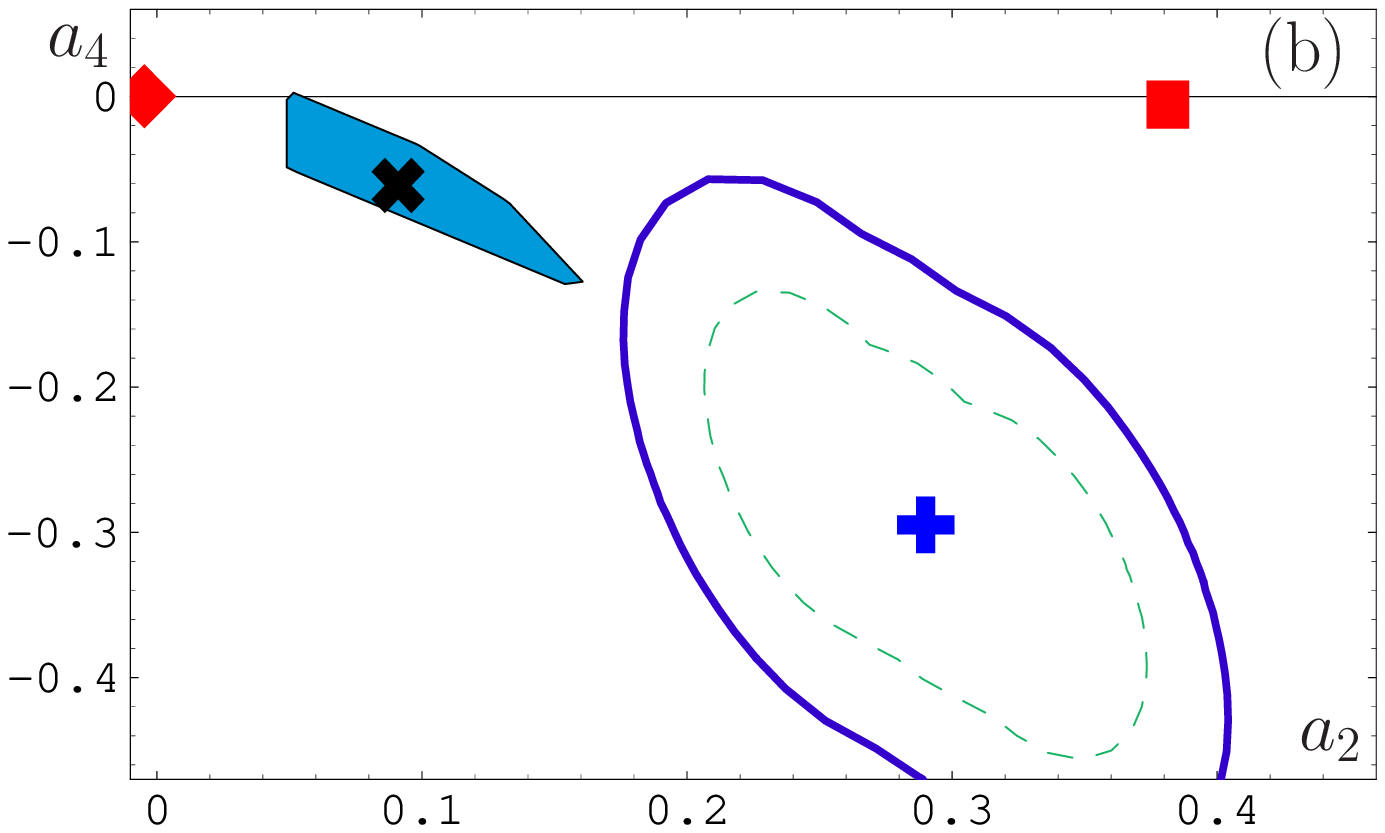}%
            ~\includegraphics[width=0.32\textwidth]{
             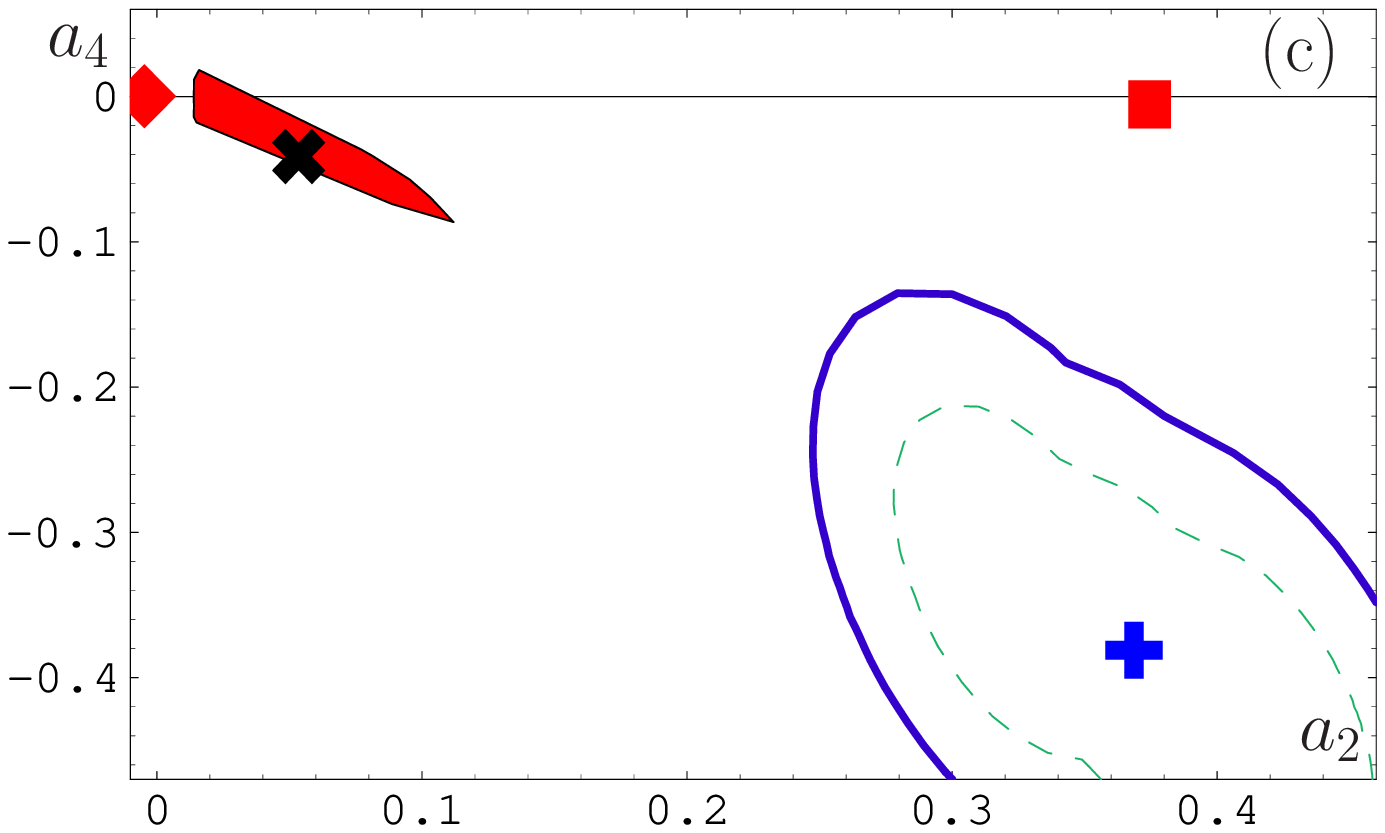}\vspace*{-3mm}}%
  \caption{Three $2\sigma$- and $1\sigma$-contours (solid and dashed lines, correspondingly) 
   of the admissible regions following from the analysis of the CLEO data 
   for different values of $\lambda^2_q$:
   (a) $\lambda^2_q=0.4~\text{GeV}^{2}$;
   (b) $\lambda^2_q=0.5~\text{GeV}^{2}$;
   (c) $\lambda^2_q=0.6~\text{GeV}^{2}$.
   \label{fig:cleo.456}\vspace*{-1mm}}
\end{figure}
The results of our analysis in~\cite{BMS02} 
are displayed in Fig.\ \ref{fig:cleo.456}.
Solid lines in all figures enclose the $2\sigma$-contours,
whereas the $1\sigma$-contours are limited by dashed lines.
The three slanted and shaded rectangles represent the constraints
on ($a_2,~a_4$) posed by the QCD SRs~\cite{BMS01}
for different values of $\lambda^2_q=0.6,~0.5,~0.4$~GeV$^{2}$
(from the left to the right).
All values are evaluated at $\mu^2=2.4$~GeV$^2$ after NLO evolution.

\begin{figure}[h]
 \centerline{\includegraphics[width=0.4\textwidth]{
     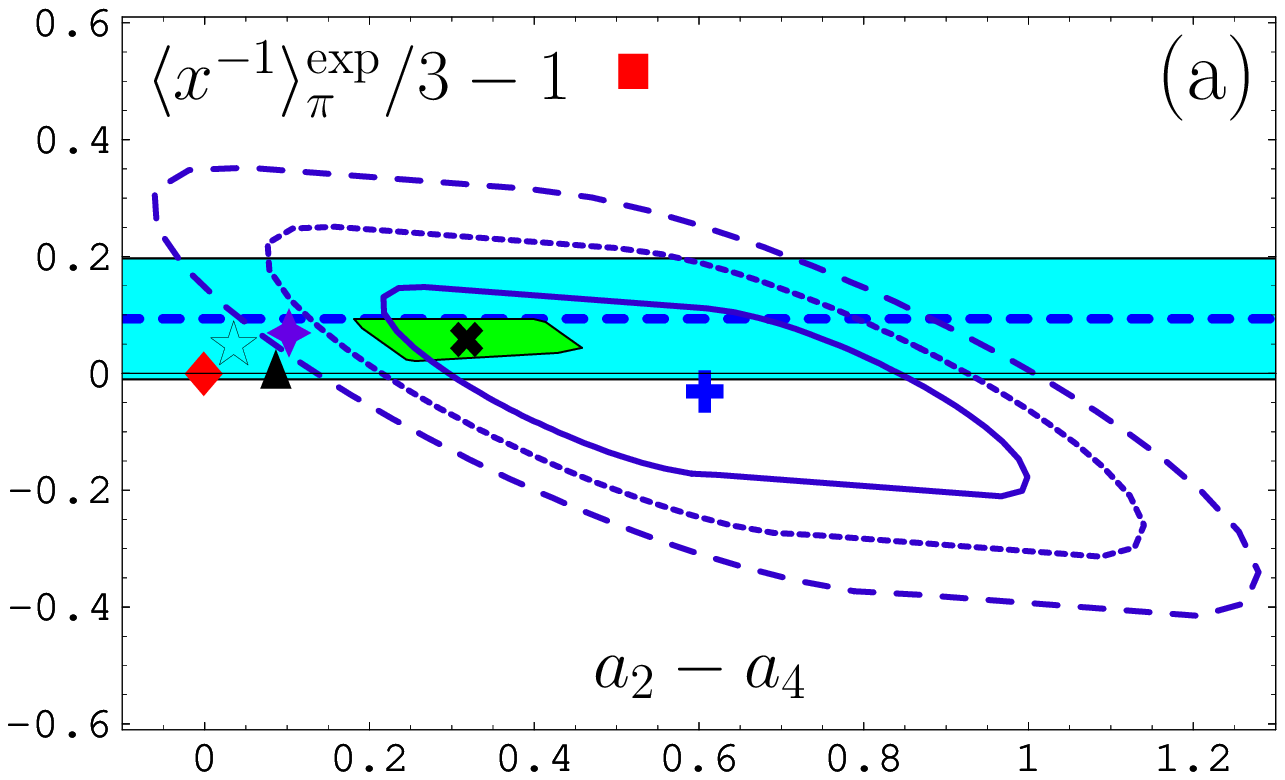}~~~~~~~~%
             \includegraphics[width=0.4\textwidth]{
     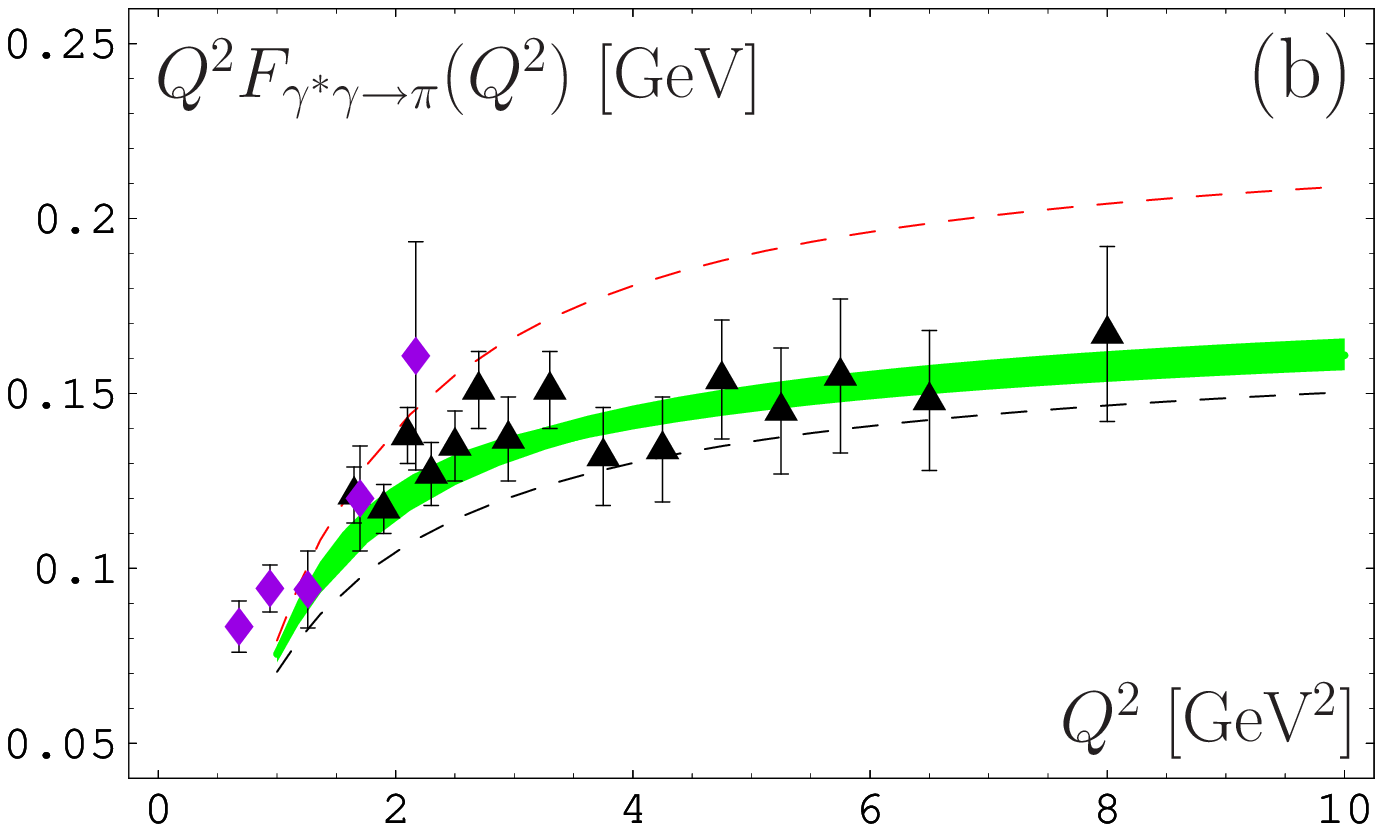}\vspace*{-3mm}}
   \caption{(a): The results of the CLEO-data analysis 
    for the pion DA parameters
    ($\displaystyle \langle x^{-1} \rangle^\text{exp}_{\pi}/3-1$,
    evaluated at $\mu^2_0 \approx 1~\text{GeV}^{2}$.
    (b): LCSR predictions for 
    $Q^2F_{\gamma^*\gamma\to\pi}(Q^2)$
    for the CZ DA (upper dashed line),
    BMS-``bunch'' (shaded strip), 
    and the asymptotic DA (lower dashed line)
    in comparison with the CELLO 
   (diamonds \protect\cite{CELLO91}) and the CLEO 
   (triangles \protect\cite{CLEO98}) experimental data,
    evaluated at $\mu^2_\text{SY}=5.76~\text{GeV}^2$ 
    with the twist-4 parameter value 
    $\delta_{\rm Tw-4}^2=0.19$~GeV$^2$~\protect\cite{BMS02}.
   \label{fig:inv.mom.ff.cello}\vspace*{-2mm}}
\end{figure}

We see that the CLEO data favor the value 
of the QCD nonlocality parameter $\lambda_q^2 = 0.4$ GeV$^2$.
We also see from Fig.\ \ref{fig:cleo.456}(c)
(and this conclusion was confirmed even 
 with a 20\% uncertainty of the twist-four magnitude---cf. 
 Fig.\ \ref{fig:inv.mom.ff.cello}(a))
that the CZ DA ({\red\footnotesize\ding{110}}) is excluded at least 
at the $4\sigma$-level, 
whereas the asymptotic DA ({\red\ding{117}}) is off at the $3\sigma$-level.
At the same time, the BMS DA (\ding{54}), 
and most of the bunch (the slanted green-shaded rectangle around the symbol \ding{54}), 
are inside the $1\sigma$-domain.
The instanton-based Bochum (\ding{73}) and Dubna ($\blacktriangle$) models 
are near but just outside the $3\sigma$-boundary
and only the Krakow model~\cite{PR01},
denoted in Fig.\ \ref{fig:inv.mom.ff.cello}(a) 
by the symbol {\violet\ding{70}},
is close to the $2\sigma$-boundary.

In Fig.\ \ref{fig:inv.mom.ff.cello}(a) 
we demonstrate the $1\sigma$-, $2\sigma$- and $3\sigma$-contours
(solid, dotted, and dashed contours around
the best-fit point ({\blue\ding{58}})), 
which have been obtained for values of the twist-4 scale parameter 
$\delta_\text{Tw-4}^2=[0.15-0.23]~\text{GeV}^{2}$.
As one sees from the blue dashed line within the hatched band
(corresponding in this figure to the mean value of 
$\displaystyle \langle x^{-1} \rangle^\text{SR}_{\pi}/3-1$
and its error bars)
the nonlocal QCD sum-rules result with its error bars
appears to be in good agreement with the CLEO-constraints 
on $\displaystyle \langle x^{-1} \rangle^\text{exp}_{\pi}$ 
at the $1\sigma$-level.
Moreover, the estimate $\displaystyle \langle x^{-1} \rangle^\text{SR}_{\pi}$
is close to $\displaystyle \langle x^{-1} \rangle^\text{EM}_{\pi}/3-1=0.24\pm 0.16$,
obtained in the data analysis of the pion electromagnetic form factor
within the framework of a different LCSR method in \cite{BKM00,BK02}.
These three independent estimates are in good agreement to each other,
giving firm support that the CLEO-data processing, on the one hand, 
and the theoretical calculations, on the other, 
are mutually consistent.

Another possibility, suggested in~\cite{Ag05b}, 
to obtain constraints on the pion DA in the LCSR analysis 
of the CLEO data would be 
to use for the twist-4 contribution a renormalon-based model
and to relate it to the parameters $a_2$ and $a_4$ of the pion DA.
Using this method, we obtained in~\cite{BMS05lat} 
renormalon-based constraints 
for the parameters $a_2$ and $a_4$
as shown in Fig.\ \ref{fig:Lat.Ren.CLEO} 
in the form of a $1\sigma$-ellipse (dashed contour).

\section{Dijet E791 data, pion form factor and CEBAF data}
\begin{figure}[h]
 \centerline{\includegraphics[width=0.85\textwidth]{
   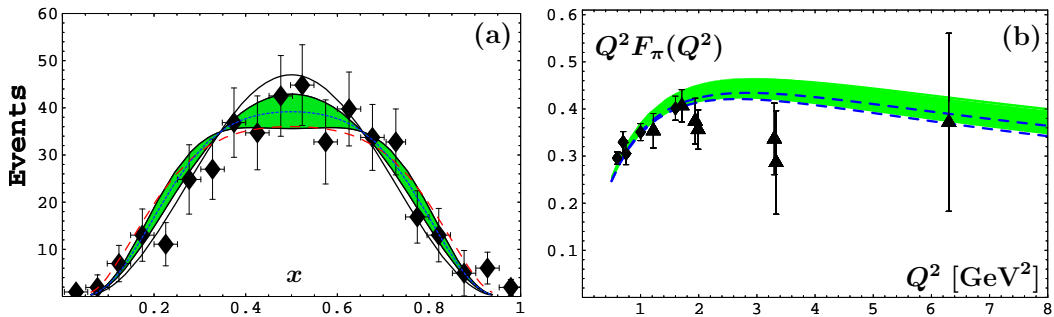}\vspace*{-3mm}}
   \caption{(a): Comparison with the E791 data
     on the diffractive dijet production of the BMS ``bunch''
     (shaded strip), the asymptotic DA (solid line), and the CZ
     (dashed line) model, using the convolution approach of~\cite{BISS02}.
    (b): The scaled pion form factor 
    calculated with the BMS\ bunch (shaded strip) and the asymptotic DA 
    (dashed lines) including nonperturbative uncertainties 
    from NLC QCD SRs~\protect{\cite{BMS01,BP06}}
    and the renormalization scheme and scale ambiguity 
    at the $O(\alpha_s^2)$ level~\cite{BPSS04}.
    The experimental data are taken from \protect{\cite{JLAB00}}
    (diamonds) and \cite{FFPI73-76} (triangles).
    \label{fig:dijet.FF.Pi}\vspace*{-2mm}}
\end{figure}

Our findings are further confirmed by the E791 data~\cite{E79102} 
on diffractive dijet $\pi+A$-production. 
This is illustrated in Fig.\ \ref{fig:dijet.FF.Pi}(a).
The main conclusion here is that all considered pion DAs
are consistent with these data, 
with a slight  preference for the BMS DA.
Indeed, following  the convolution procedure of~\cite{BISS02},
we found~\cite{BMS02} the following values of $\chi^2$ 
for the three types of pion DAs:
12.56 (asymptotic DA), 10.96 (BMS bunch), and 14.15 (CZ DA).

It is worth mentioning the results of our analysis of
the pion electromagnetic form factor
using the NLC-based pion DA and analytic perturbative QCD~\cite{BPSS04}.
These results are in excellent agreement
with the CEBAF data on the pion form factor, 
as illustrated in Fig.\ \ref{fig:dijet.FF.Pi}(b),
where the green strip includes both the NLC QCD SR uncertainties,
generated by the bunch of the allowed pion DAs,
and by the scale-setting ambiguities at the NLO level of QCD perturbation theory.

From the phenomenological point of view, 
the most interesting result here 
is that the BMS pion DA~\cite{BMS01} 
(out of the ``bunch'' of similar doubly-peaked endpoint-suppressed pion DAs) 
yields predictions for the electromagnetic form factor of the pion,
which are very close to those obtained with the asymptotic pion DA.
Conversely, we see from Fig.\ \ref{fig:dijet.FF.Pi}(b) 
that a small deviation 
of the prediction
for the pion form factor 
from that obtained with the asymptotic pion DA
(dashed lines)
does not necessarily imply that the underlying pion DA 
has to be close
to the asymptotic profile.
Much more important is the behavior of the pion DA in the endpoint
region $x\to 0\,, 1$.
For more details, we refer to~\cite{BPSS04}.

\section{New lattice data and pion DA}
Rather recently, new high-precision lattice measurements 
of the second moment of the pion DA
$\langle{\xi^2}\rangle_{\pi} = \int_0^1(2x-1)^2\varphi_\pi(x)\,dx$
appeared~\cite{DelD05,Lat06}.
Both cited groups extracted from their respective simulations 
values of $a_2$ at the Schmedding--Yakovlev scale
$\mu^2_\text{SY}$ around $0.24$,
but with different error bars.

It is remarkable that these lattice results are in striking agreement
with the estimates of $a_2$ from both the NLC QCD SRs~\cite{BMS01} 
and also from the CLEO-data analyses---based 
on LCSR---\cite{SY99,BMS02}, 
as illustrated in Fig.\ \ref{fig:Lat.Ren.CLEO}(a), 
where the lattice results of~\cite{Lat06}
are shown in the form of a vertical strip, 
containing the central value with associated errors.
Remarkably, the value of $a_2$ of the displayed lattice measurements
(middle line of the strip) is very close to the CLEO best fit 
in~\cite{BMS02} ({\blue\ding{58}}).

\section{Improved model for NLCs and pion DA}
The quark-gluon-antiquark condensates are usually parameterized
in the following form
\begin{eqnarray}
\langle{\bar{q}(0)\gamma_\mu\left(-g\sum\limits_{a}t^a A^a_\nu(y)\right)q(x)}\rangle
  &=&(y_\mu x_\nu-g_{\mu\nu}(yx))\overline{M}_1(x^2,y^2,(y-x)^2)\nonumber\\
  &+&
      (y_\mu y_\nu-g_{\mu\nu}y^2)\overline{M}_2(x^2,y^2,(y-x)^2)\,,\nonumber
\\
\langle{\bar{q}(0)\gamma_5\gamma_\mu\left(-g\sum\limits_{a}t^a A^a_\nu(y)\right)q(x)}\rangle
  &=&
       i\varepsilon_{\mu\nu yx}\overline{M}_3(x^2,y^2,(y-x)^2)\,,
\vspace{-5mm}\nonumber
\end{eqnarray}
where
$\displaystyle\overline{M}_i(x^2,y^2,z^2)\!=\!
     A_i\int\!\!\!\!\int\limits_{\!0}^{\,\infty}\!\!\!\!\int\!\!
        d\alpha \, d\beta \, d\gamma \,
         f_i(\alpha ,\beta ,\gamma )\,
          e^{\left(\alpha x^2+\beta y^2+\gamma z^2\right)/4}$
with
$A_{1,2,3}=A_0\left(-\frac32,2,\frac32\right)$.
The minimal model of the nonlocal QCD vacuum suggests 
the following Ansatz
\begin{eqnarray}
 \label{eq:Min.Anz.qGq}
  f_i^\text{min}\left(\alpha,\beta,\gamma\right)
   = \delta\left(\alpha -\Lambda\right)\,
       \delta\left(\beta -\Lambda\right)\,
        \delta\left(\gamma -\Lambda\right)
\end{eqnarray}
with $\Lambda=\frac12\lambda_q^2$ and faces problems 
with the QCD equations of motion 
and the gauge invariance of the 2-point correlator of vector currents.
In order to fulfil the QCD equations of motion exactly
and minimize the non-transversity of $V-V$ correlator,
an improved version of the QCD vacuum model was suggested in~\cite{BP06} 
with 
\begin{eqnarray}
 \label{eq:Imp.Anz.qGq}
 f^\text{imp}_i\left(\alpha,\beta,\gamma\right)
  = \left(1 + X_{i}\partial_{x} + Y_{i}\partial_{y} + Y_{i}\partial_{z}\right)
         \delta\left(\alpha-x\Lambda\right)
          \delta\left(\beta-y\Lambda\right)
           \delta\left(\gamma-z\Lambda\right)\,,
\end{eqnarray}
where $\Lambda=\frac12\lambda_q^2$ and
\begin{subequations}
\begin{eqnarray}
  X_1 &=& +0.082\,;~X_2 = -1.298\,;~X_3 = +1.775\,;~x=0.788\,;~~~\\
  Y_1 &=& -2.243\,;~Y_2 = -0.239\,;~Y_3 = -3.166\,;~y=z=0.212\,.~~~
\end{eqnarray}
\end{subequations}
\begin{figure}[t]
 \centerline{\includegraphics[width=0.4\textwidth]{
     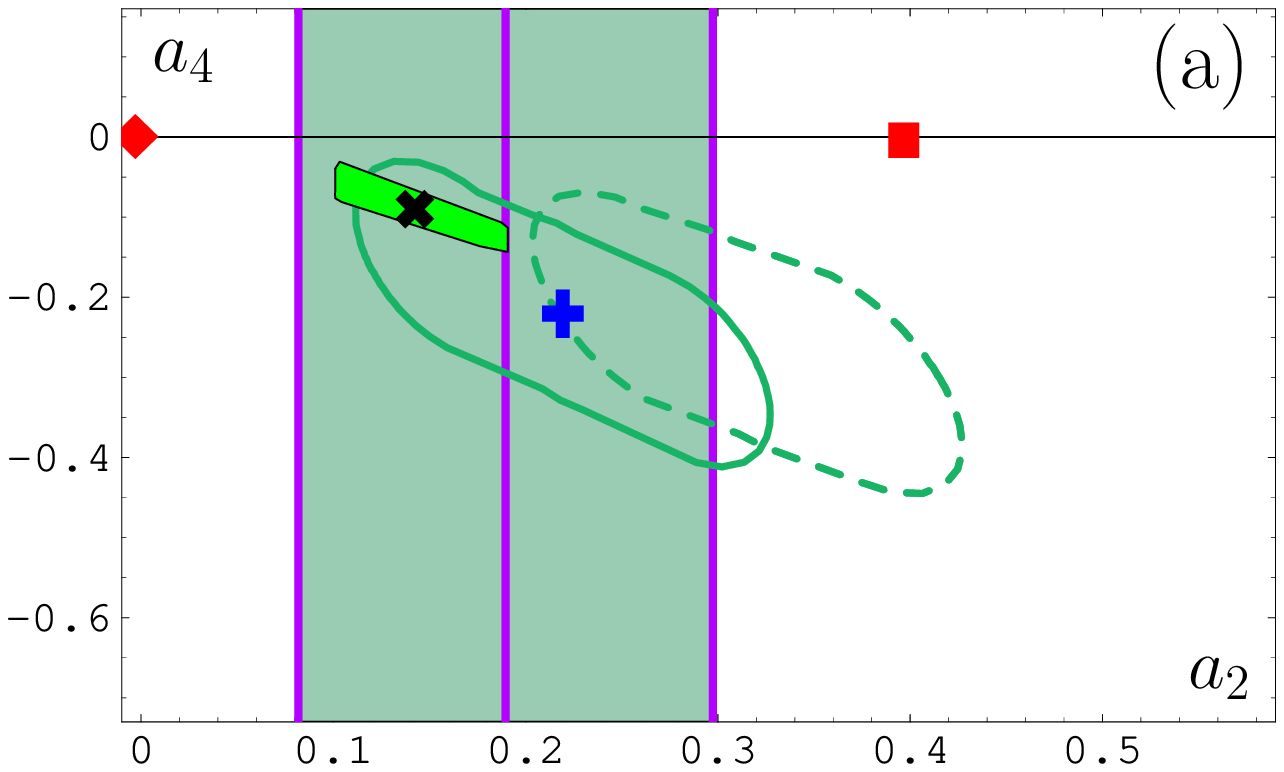}~~~~~~~
             \includegraphics[width=0.4\textwidth]{
     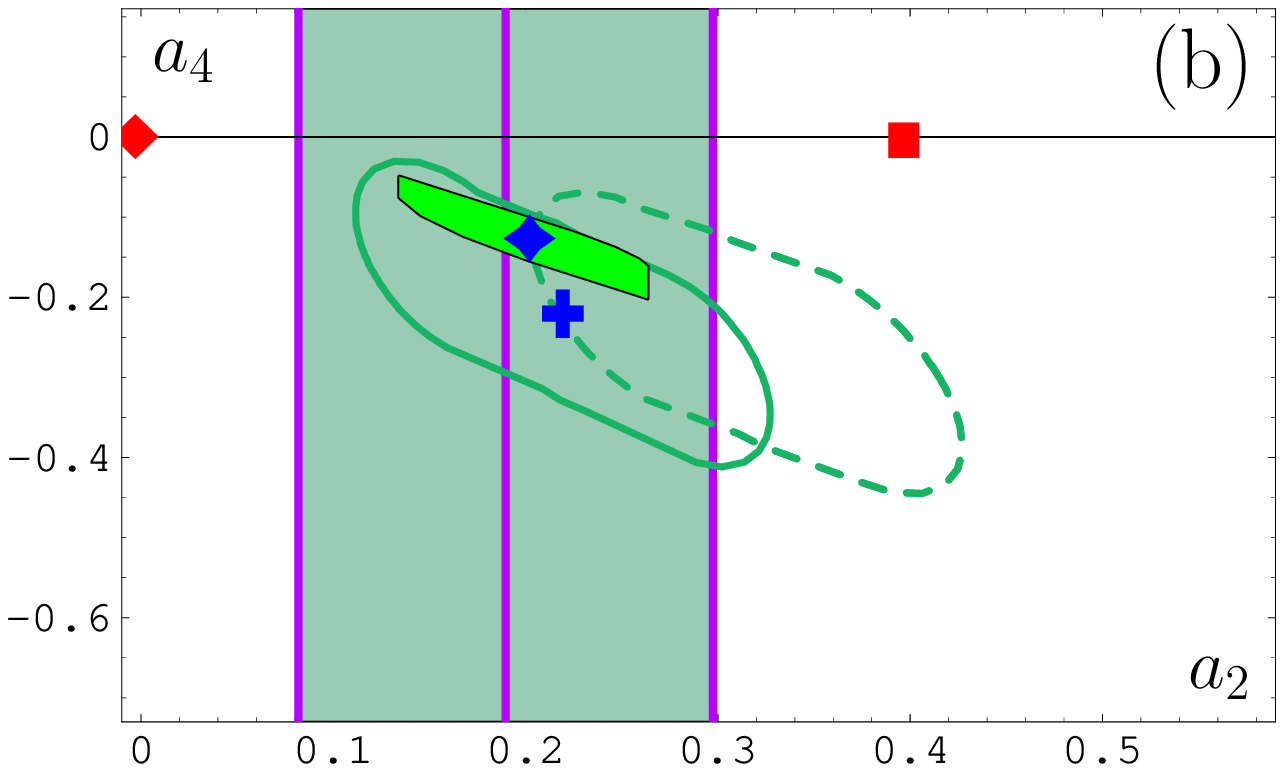}\vspace*{-3mm}}
       \caption{The results of the CLEO-data analysis for the pion DA
        parameters $a_2$ and $a_4$, evaluated at $\mu^2_\text{SY}=5.76~\text{GeV}^2$,
        are shown. The lattice results of \protect\cite{Lat06} are displayed 
        as a shaded area, whereas the renormalon-based $1\sigma$-ellipse 
        of~\protect\cite{BMS05lat} is denoted by the green dashed line.
        The small slanted rectangle on panel (a) represents the corresponding results
        for the BMS-``bunch'', whereas the modified, somewhat larger, slanted rectangle
        corresponding to the improved Gaussian model of QCD vacuum,
        is shown on panel (b).
        \label{fig:Lat.Ren.CLEO}\vspace*{-1mm}}
\end{figure}
Then, the NLC sum rules~(\ref{eq:NLC.SR.pion.DA}) gives rise to
a modified ``bunch'' of two-parameter pion DA models (\ref{eq:pi.DA.2Geg})
at $\mu^2=1.35$ GeV$^2$~\cite{BP06}.
The coordinates of the central point {\blue\ding{70}} are 
$a_2=0.268$ and $a_4=-0.186$. 
These values correspond to $\langle{x^{-1}}\rangle_\pi^{\text{bunch}} = 3.24\pm0.20$,
which is in agreement with the result provided by an independent sum rule, 
viz.,
$\langle{x^{-1}}\rangle_\pi^{\text{SR}}=3.40\pm0.34$.
The allowed values of both bunch parameters $a_2$ and $a_4$
after NLO-evolution to $\mu^2=5.76$ GeV$^2$ 
are shown in Fig.\ \ref{fig:Lat.Ren.CLEO} in the form 
of shaded slanted rectangles around the central points \ding{54}
and {\blue\ding{70}}.

\begin{figure}[b]
 \centerline{\includegraphics[width=0.4\textwidth]{
     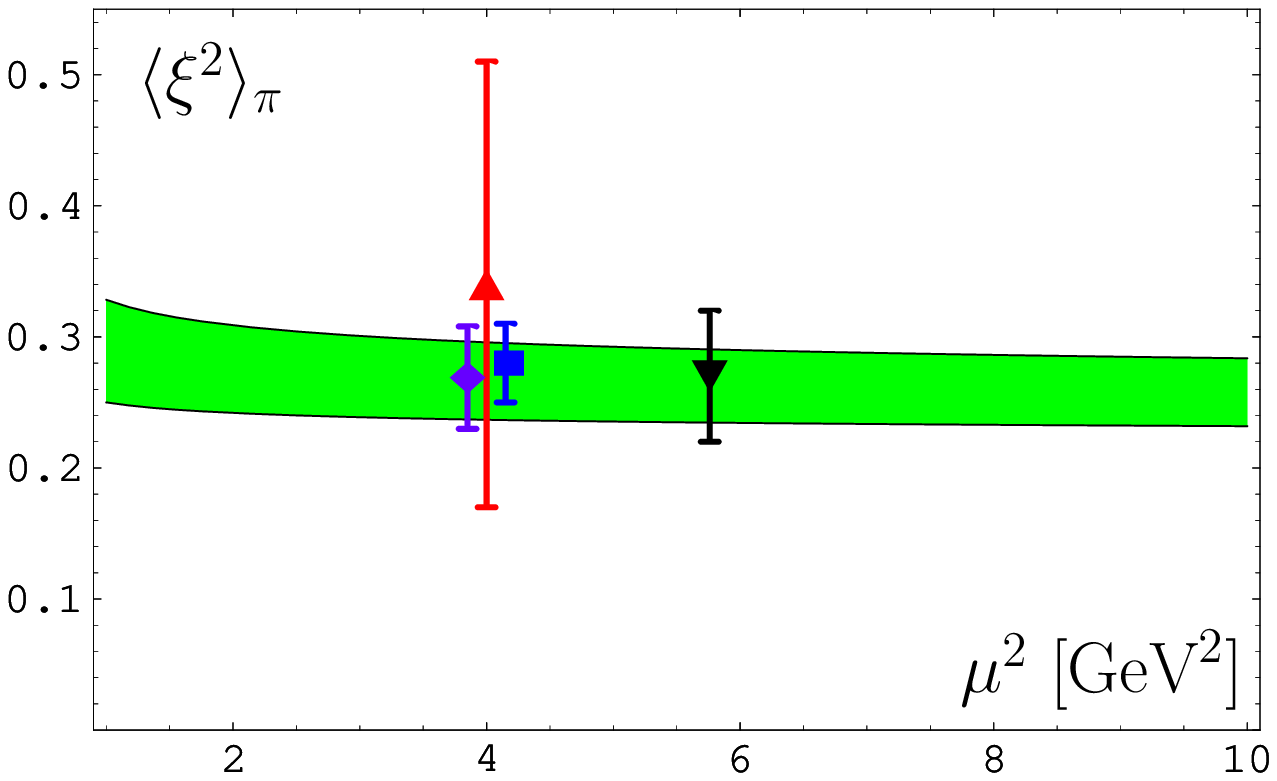}~~~~~~~
             \includegraphics[width=0.409\textwidth]{
     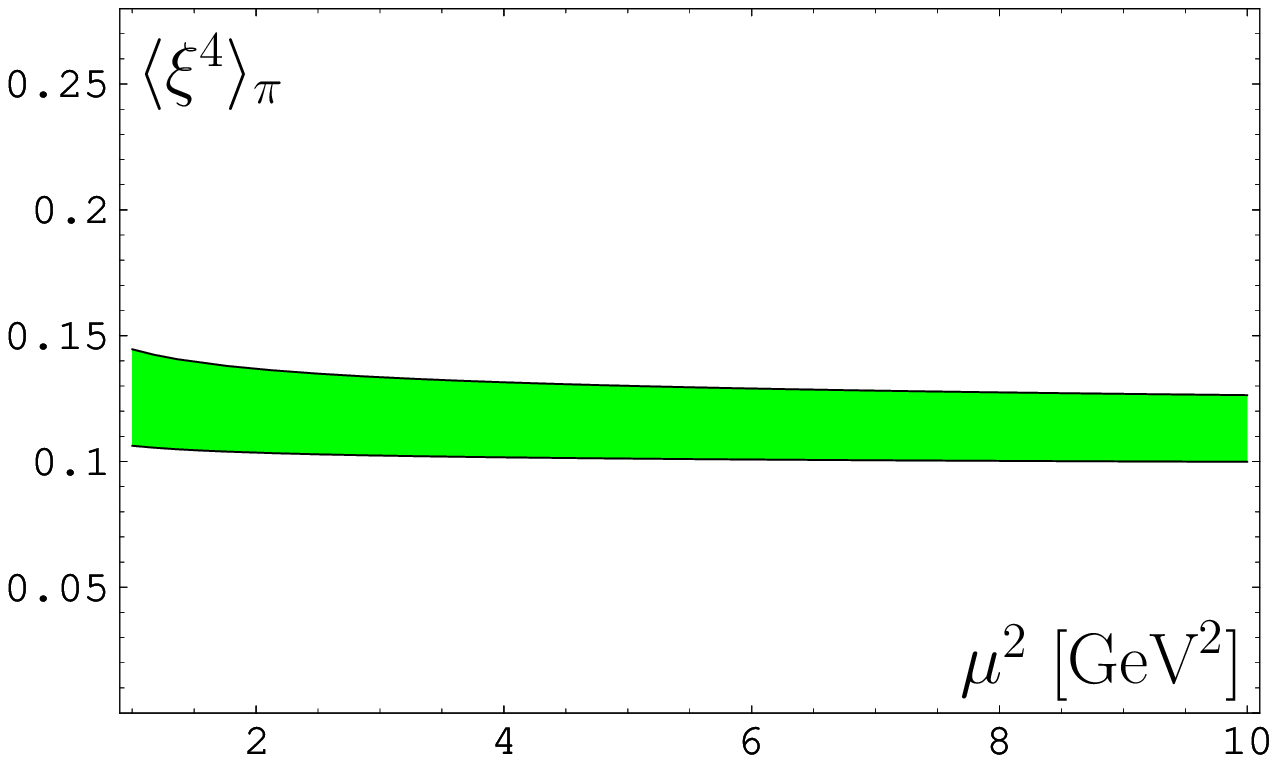}\vspace*{-3mm}}
  \caption{Evolution of $\langle{\xi^2}\rangle_\pi(\mu^2)$ (left panel) and
        $\langle{\xi^4}\rangle_\pi(\mu^2)$ (right panel) with $\mu^2\in1-10$~GeV$^2$.
        The green strip on both panels corresponds to the unified results 
        of the QCD Sum Rules with NLCs (minimal and improved models of the QCD vacuum).
        On the left panel, we also show the lattice results with their corresponding 
        error-bars:
        {\small\red\ding{115}} \protect\cite{MS88},
        {\small\ding{116}} \protect\cite{DelD05},
        {\violet\ding{117}} \protect\cite{Lat06},
        {\small\blue\ding{110}} \protect\cite{Lat07}.
        \label{fig:Mom24.Evol.Lat}  }
\end{figure}
We emphasize in this context that the BMS model~\cite{BMS01}, 
shown in Fig.\ \ref{fig:Lat.Ren.CLEO}(a) by the symbol {\ding{54}},
is inside the allowed region extracted from the improved QCD vacuum model.
This means that all the characteristic features of the original BMS bunch 
are also valid for the improved bunch.
Again the NLC-based model DAs are end-point suppressed,
though they are doubly humped.

We see from Fig.\ \ref{fig:Lat.Ren.CLEO}(b) 
that the improved bunch~\cite{BP06} 
appears to be in somewhat better agreement 
with the recent lattice results~\cite{Lat06},
shown in the form of a vertical strip, 
containing the central value with its associated errors.
Remarkably, the value of $a_2$ of the displayed lattice measurements
(middle line of the strip) is very close to the central point 
of the improved bunch ({\blue\ding{70}}), 
whereas the whole admissible region (slanted rectangle)~\cite{BP06}
is inside the strip\footnote{%
This statement is valid for the BMS ``bunch'' as well, 
as can be seen from Fig.\ \ref{fig:Lat.Ren.CLEO}(a).}
and also inside the standard CLEO $1\sigma$-ellipse.
Furthermore we see from this figure 
that the higher the precision of the lattice simulations,
the closer they are to the results for $\langle{\xi^2}\rangle_\pi(\mu^2)$ 
of the NLC-based QCD SRs.
It remains to be seen 
whether this agreement will persist also for the $\langle{\xi^4}\rangle_\pi(\mu^2)$
moment,
once it is calculation will become possible on the lattice~\cite{BraM07}.

\section{Pion DA and Drell--Yan $\pi N$ process}
\begin{figure}[h]
 \begin{minipage}{\textwidth}
  \begin{minipage}{0.49\textwidth}%
   \centerline{\includegraphics[width=\textwidth]{
    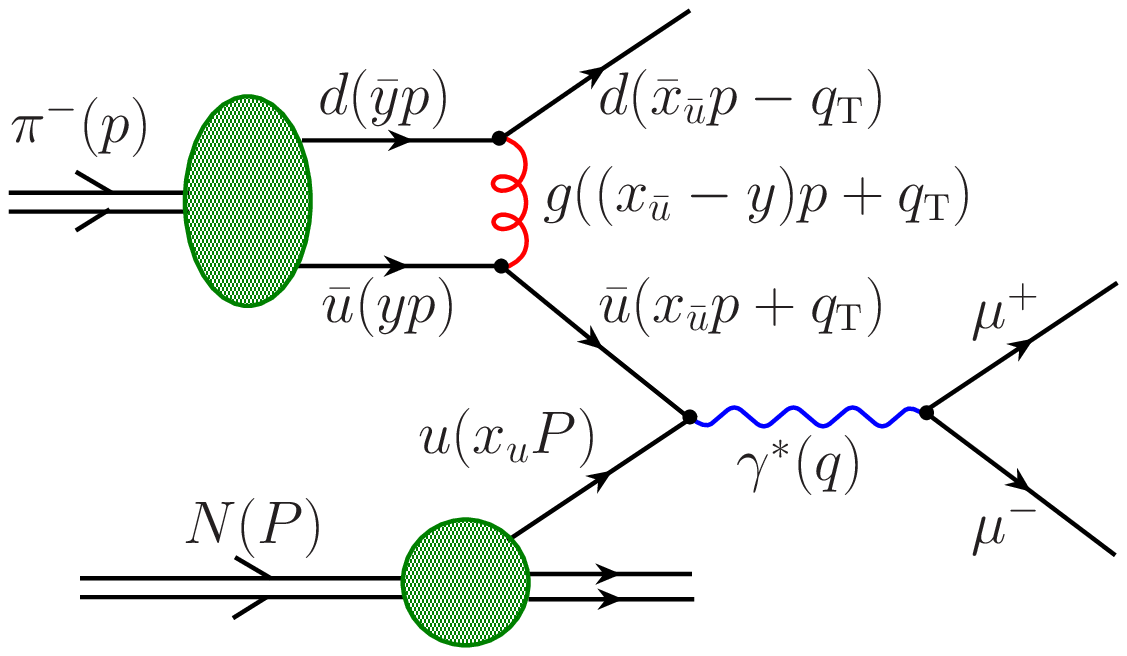}}%
  \end{minipage}~~~
  \begin{minipage}{0.49\textwidth}%
   \vspace*{-5mm}
 
   \centerline{\includegraphics[width=\textwidth]{
    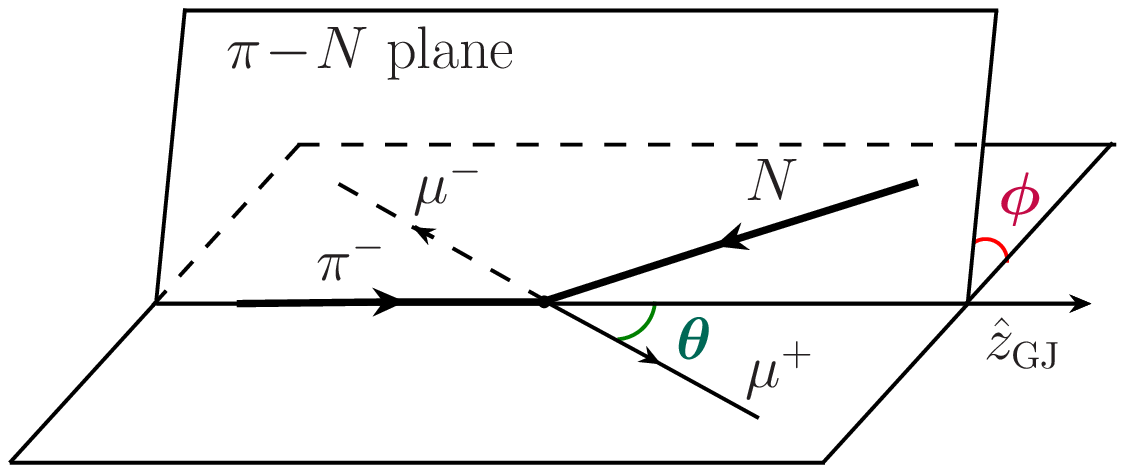}}
  \end{minipage}
  \caption{Left panel:
   Graphical representation of the Drell--Yan
   process $\pi^{-}N\to \mu^{+}\mu^{-}X$.
   Right panel: Angular definitions of the Drell--Yan 
   process in the center of mass frame of the produced massive lepton 
   pair. The axis $\hat{z}_{\rm GJ}$ denotes the pion direction in the
   Gottfried--Jackson (GJ) frame.
   \label{fig:dy-kinem}\vspace*{-2mm}}
 \end{minipage}
\end{figure}

The DY process is the dominant mechanism to produce lepton pairs
with a large invariant mass $Q^2$ in hadronic collisions, like
$\pi^{\pm}N$ scattering.
In this model a massive muon pair is created
through the electromagnetic annihilation of an antiquark from the beam
pion and a quark from the nucleon target, 
as depicted on the left panel of Fig.\ \ref{fig:dy-kinem}.

Here $p_{\bar{u}}=x_{\bar{u}}P$ is the momentum of the annihilating antiquark 
from the pion.
Typical values of the kinematical parameters
$s=(p_\pi+P_N)^2$,
$Q^2=q^2$,  and $Q_{T}^2=-q_{T}^2$, see Fig.\ \ref{fig:dy-kinem},
in the case 
of the FNAL experiment E615 are:
$s=500~\text{GeV}^2$, 
$Q^2=16-70~\text{GeV}^2$, and
$\rho\equiv Q_{T}/Q=0-0.5$.
\begin{figure}[t]
 \centerline{\includegraphics[width=0.31\textwidth]{
    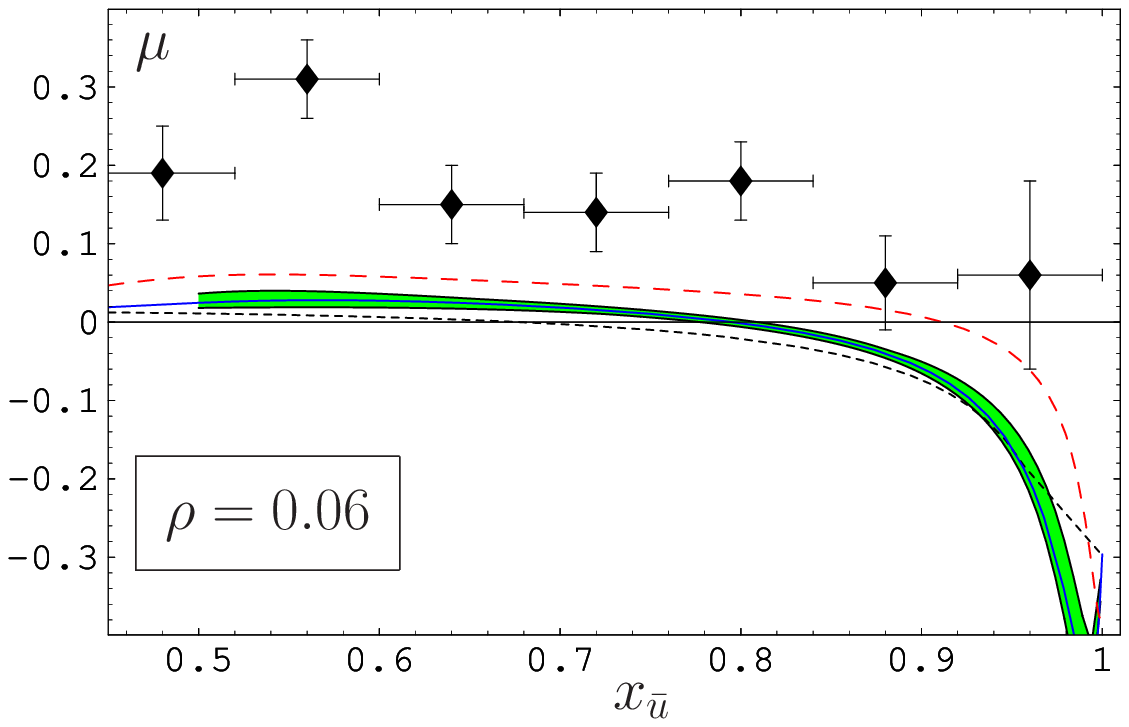}~~~%
   \includegraphics[width=0.31\textwidth]{
    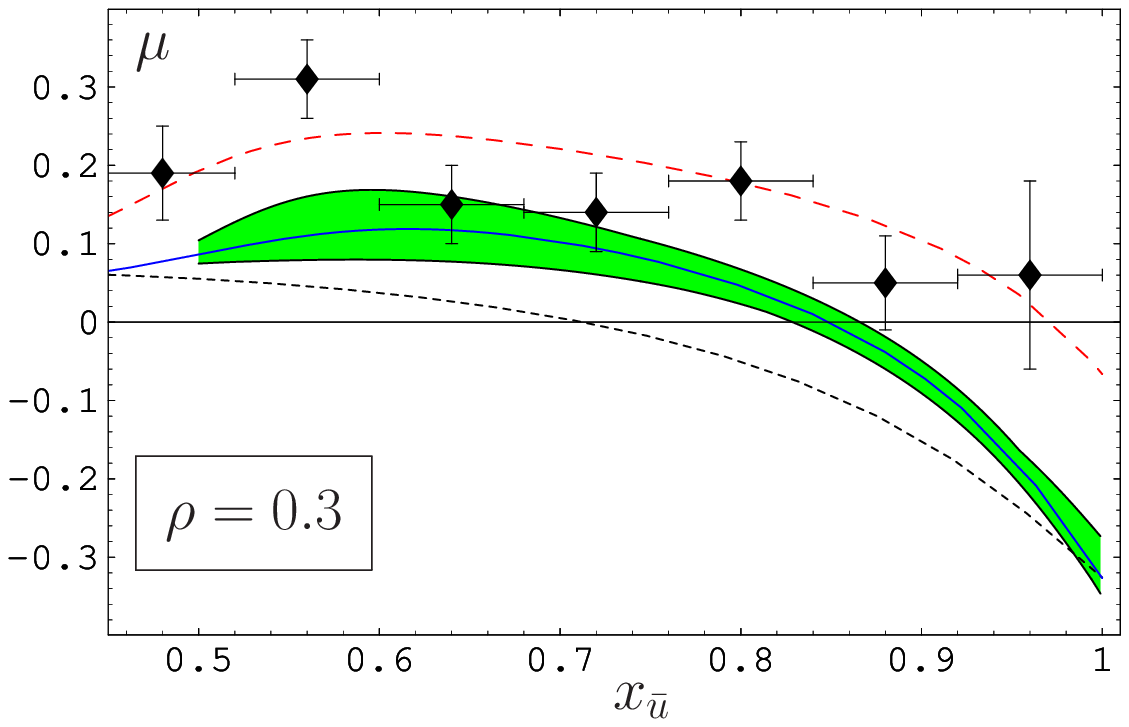}~~~%
   \includegraphics[width=0.31\textwidth]{
    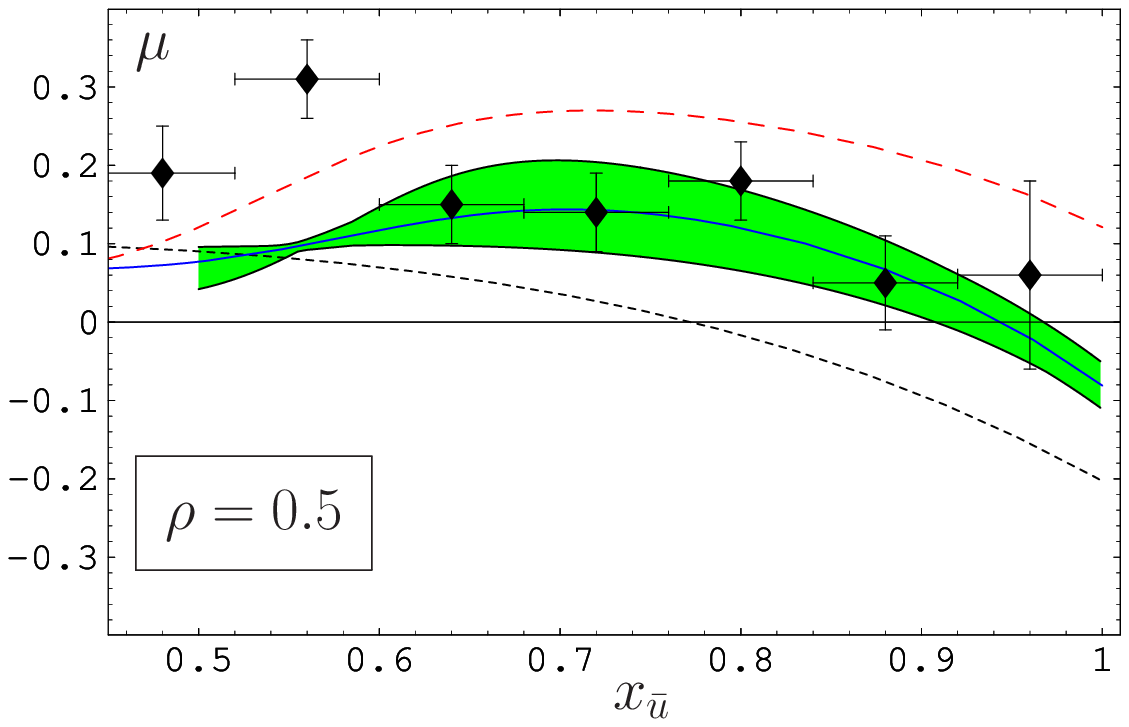}\vspace*{-3mm}}
   \caption{Results for the angular distribution
   parameter $\mu$ as a function of
   $x_{\bar{u}}\equiv x_{\pi}$ for different values of
   $\rho\equiv Q_{T}/Q$.
   The green strip corresponds to the BMS bunch of the pion DAs~\protect\cite{BMS01}
   with the solid line representing the BMS DA, 
   while the dotted solid line shows the result for the asymptotic DA, 
   and the dashed line denotes the prediction for the endpoint-dominated
   CZ DA. The data are taken from \protect\cite{Con89}.
   \label{fig:unpol-angular}\vspace*{-2mm}}
\end{figure}
As $x_{\bar{u}}\to 1$,
$p_{\bar{u}}^2$ becomes large and far spacelike 
and, therefore, 
it is sufficient to consider the $u$-quark 
to be nearly free and on-shell:
$x_{u}=x_{N}$ (no transverse momenta).
On the right panel of Fig.\ \ref{fig:dy-kinem}
we show the angular-distribution parameters 
$\theta$, the polar angle measuring the $\mu^+$ direction 
in the Gottfried--Jackson system of axes, 
and $\phi$, the azimuthal angle between the $\pi^-\mu^+\mu^-$ 
and $\pi^{-}N$ planes in the lepton rest frame.

\begin{figure}[hb]
 \centerline{\includegraphics[width=0.30\textwidth]{
  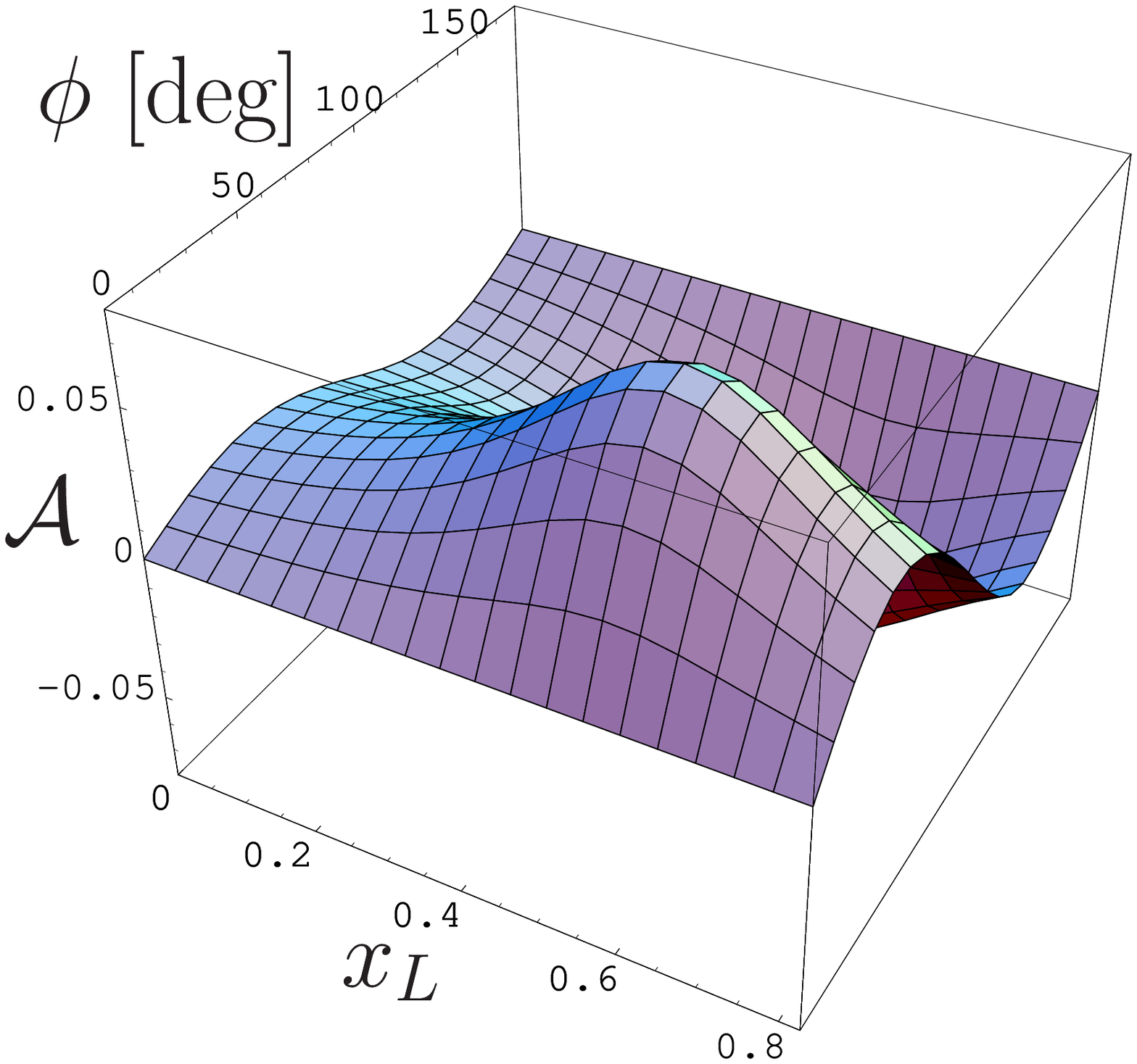}~~~%
   \includegraphics[width=0.30\textwidth]{
  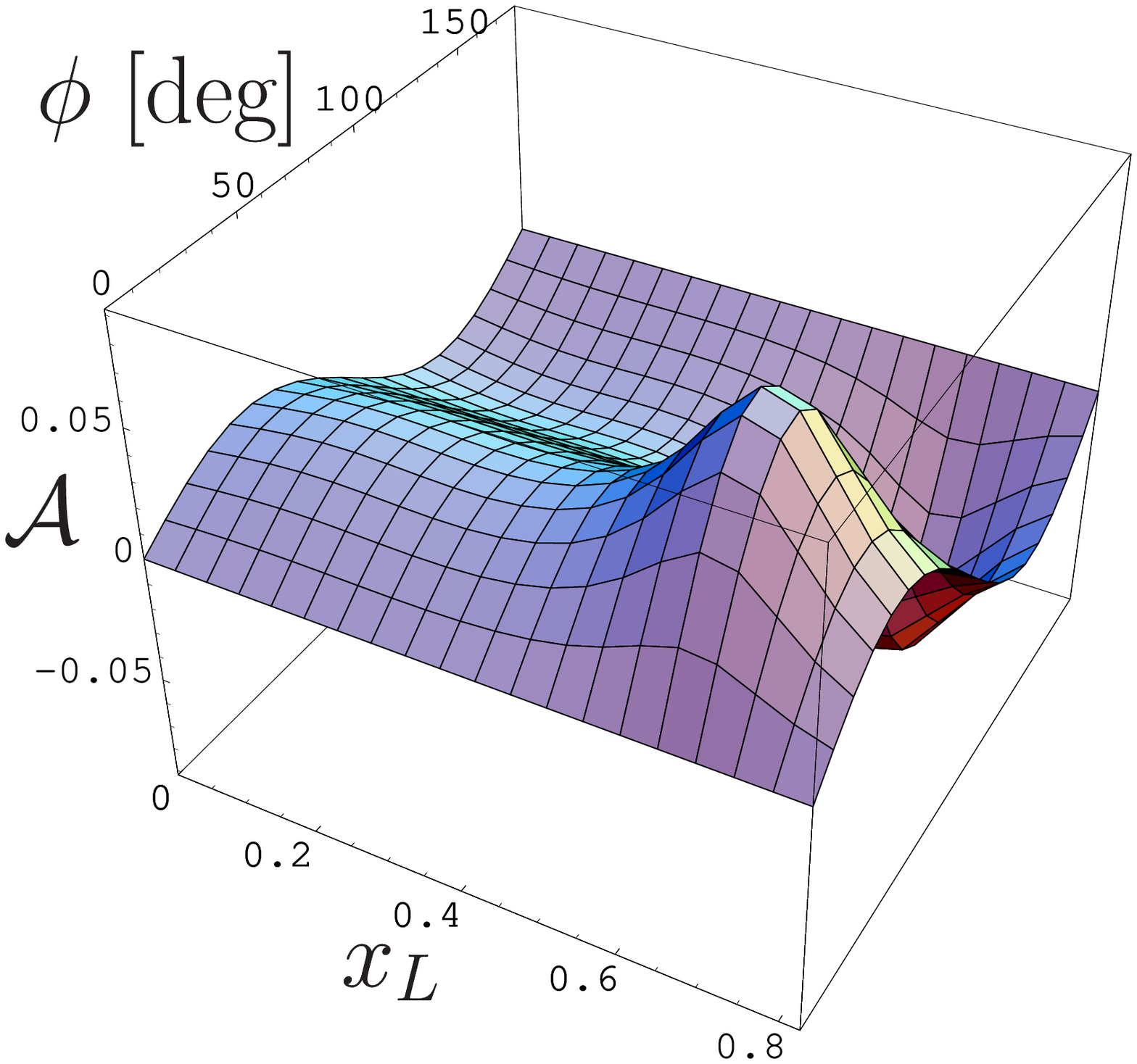}~~~%
   \includegraphics[width=0.30\textwidth]{
  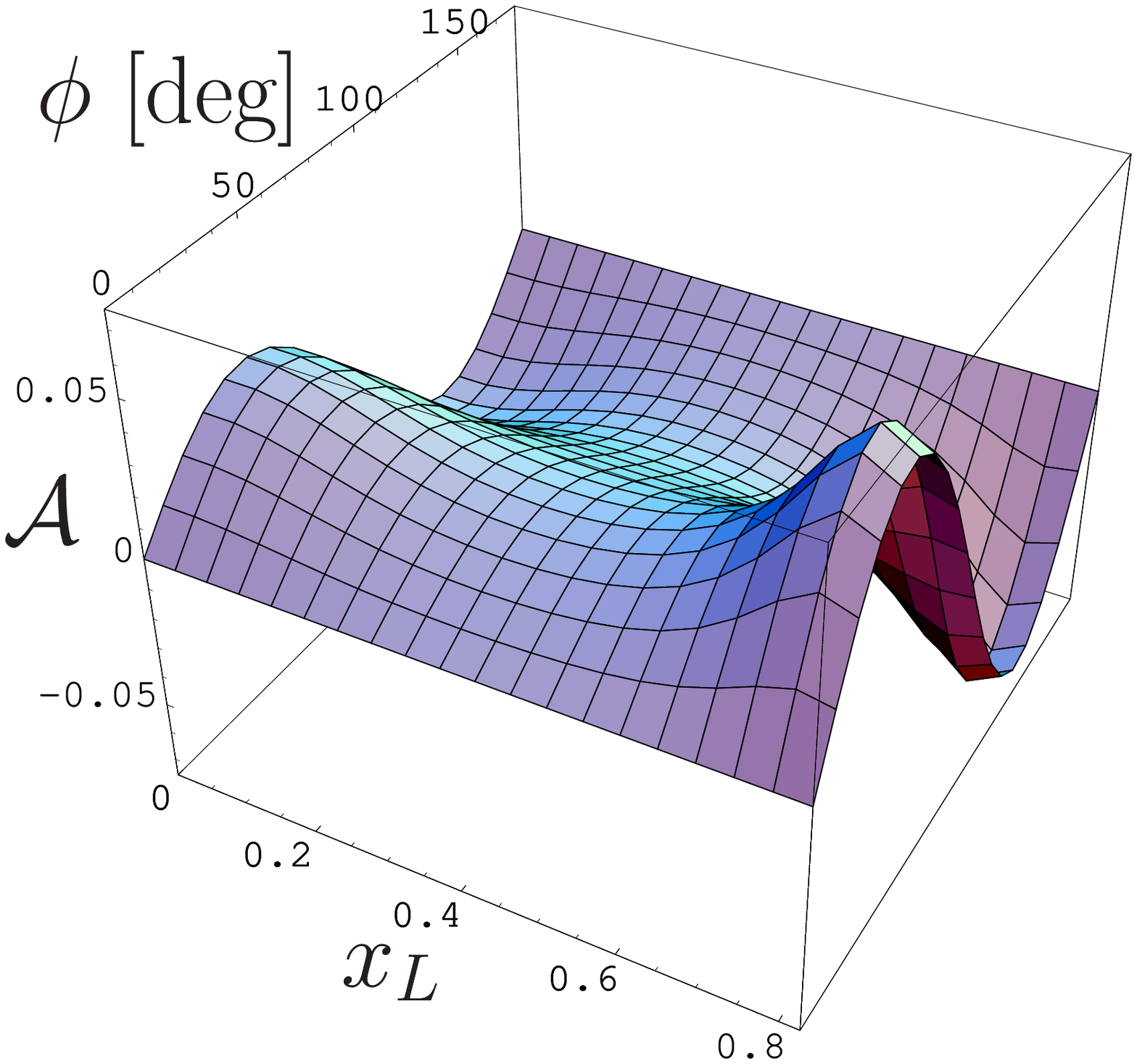}\vspace*{-3mm}}
   \caption{3D-plots of the azimuthal asymmetry
   ${\cal A}(\phi,x_L,\rho=0.3)$ at $q^2=16~\text{GeV}^2$ 
    and $s = 100~\text{GeV}^2$
    for three different choices of the pion DA:
    Asymptotic (left), BMS model (center), and CZ model (right).
   \label{fig:rho-evo-Asy-030-3D}\vspace*{-3mm}}
\end{figure}

For the DY reaction with an unpolarized target,
the angular distribution of the $\mu^{+}$ in the pair rest frame
can be written in terms of the kinematic variables $\lambda, \mu, \nu$
as follows~\cite{BBKM94}:
\begin{eqnarray}
  \frac{d^{5}\sigma(\pi^{-}N
  \to\mu^{+}\mu^{-}X)}{dQ^{2}dQ_{T}^{2}dx_{L}\,d\cos\theta d\phi}
  \propto N(\tilde{x},\rho)
           \Big(1 
              + \lambda \cos^{2}\theta 
              + \mu \sin 2\theta \cos \phi 
              + \frac{\nu}{2} \sin^{2}\theta \cos 2\phi
           \Big)\,.
\label{eq:dif-cross}
\end{eqnarray}
Adopting this convolution procedure,
we found~\cite{BST07} 
the results
presented in Fig.\ \ref{fig:unpol-angular}:
We see that the agreement of the chosen pion DA model with the unpolarized
E615 (FNAL) data depends on the value of the parameter $\rho$.
It seems that these data cannot make a clear distinction in favor
of one particular pion DA.
On the other hand, for the asymmetry of the polarized DY $\pi^{-}N$ process
we found (using the convolution procedure of~\cite{BMT95}) 
the results displayed in Fig.\ \ref{fig:rho-evo-Asy-030-3D}.
We may come to the conclusion
that the asymmetry ${\cal A}(\phi,x_L)$ 
can be used to discriminate different proposed pion DA models,
provided the value of $\rho$ can be fixed by experiment~\cite{BST07}.

\section{Conclusions}
Let us conclude with the following observations:\\
(i) The NLC QCD SR method for the pion DA gives us 
admissible bunches of pion DAs for each value of $\lambda_q^2\gtrsim0.4$~GeV$^2$.
(ii) The NLO LCSR method produces new constraints
on the pion DA parameters ($a_2$ and $a_4$)
in conjunction with the CLEO data.
(iii) Comparing the results of the NLC SRs
with the new CLEO-data constraints,
allows us to fix the value of the QCD vacuum nonlocality
to be $\lambda_q^2 \simeq 0.4~$GeV$^{2}$.
(iv) The bunch of pion DAs from NLC QCD SRs agrees well 
with the E791-data  on the diffractive dijet $\pi+A$-production, 
with the JLab F(pi) data on the pion electromagnetic form factor, 
and with the latest high-precision lattice data.
(v) Taking into account the QCD equations of motion
 for the NLCs and the transversity of the vacuum polarization,
 allows us to shift the pion DA bunch just inside the $1\sigma$-ellipse
 of the CLEO-data constraints.
(vi) In the polarized Drell--Yan $\pi^{-}N$ process
 one might be able to discriminate among different pion DAs,
 once the value of $\rho$ will be known by measurements.  

\begin{acknowledgments}\vspace*{-3mm}
One of us (A.~P.~B.) would like to thank the organizers 
of the Conference ``Hadron Structure--07'' 
(Modra-Harm{\'o}nia, Slovakia, Sept. 3--7, 2007) for the invitation and support.
This work was supported in part by 
the Russian Foundation for Fundamental Research, 
grants No.\ 06-02-16215 and 07-02-91557,
the BRFBR--JINR Cooperation Programme (contract No.\ F06D-002),
the Heisenberg--Landau Programme under grant 2007, 
and the Deutsche Forschungsgemeinschaft
(project DFG 436 RUS 113/881/0).
\end{acknowledgments}



\end{document}